\documentclass[12pt]{iopart}
\usepackage[normalem]{ulem}
\usepackage[dvipsnames]{xcolor}
\usepackage{hyperref}
\usepackage[noabbrev]{cleveref} 
\usepackage{acronym}

\usepackage{cite}   

\usepackage{booktabs}

\usepackage{siunitx}
\usepackage{graphicx}
\usepackage{tikz}
\usetikzlibrary{positioning}
\usepackage{iopams}

\usepackage{etoolbox}
\newrobustcmd\ubold{\DeclareFontSeriesDefault[rm]{bf}{b}\bfseries}

    \acrodef{QQM}{quaternionic quantum mechanics}
    \acrodef{CQM}{complex quantum mechanics}
    
\begin{document}

\crefformat{equation}{#2(#1)#3}     
\crefformat{appendix}{#2#1#3}
\Crefformat{appendix}{#2#1#3}
\crefformat{subappendix}{#2#1#3}
\Crefformat{subappendix}{#2#1#3}

\newcommand{\abs}[1]{\left\vert #1\right\vert}
\newcommand{\mean}[1]{\langle #1 \rangle} 
\newcommand{\defeq}{\equiv}  
\newcommand{\scr}[1]{^\mathrm{#1}} 

\newcommand{\twoline}[3]{%
\begin{tabular}{@{}#1@{}}%
	#2 \\%
	#3 \\%
\end{tabular}}
\newcommand{\tcellc}[1]{\multicolumn{1}{c}{#1}}
\newcommand{\phle}{\phantom{<\;}}
\newcommand{\phm}{\phantom{-}}

\title{Towards probing for hypercomplex quantum mechanics in a waveguide interferometer}

\author{S Gstir$^1$, E Chan$^{1,2}$, T Eichelkraut$^3\footnote{Present address: Janova GmbH, Mühlhenstraße 96, 07745 Jena, Germany}$, A Szameit$^4$, R Keil$^1$ and G Weihs$^1$\newline
}

\address{$^1$ Institut für Experimentalphysik, Universität Innsbruck, Technikerstr. 25, 6020 Innsbruck, Austria}
\address{$^2$ École Normale Supérieure de Lyon, 65 Allée d'Italie, 69007 Lyon, France}
\address{$^3$ Institute of Applied Physics, Abbe Center of Photonics, Friedrich-Schiller-Universität Jena, Max-Wien-Platz 1, 07743 Jena, Germany}
\address{$^4$ Institut für Physik, Universität Rostock, Albert-Einstein-Str. 23, 18059 Rostock, Germany}
\ead{sebastian.gstir@uibk.ac.at}
\vspace{10pt}

\begin{abstract}
We experimentally investigate the suitability of a multi-path waveguide interferometer with mechanical shutters for performing a test for hypercomplex quantum mechanics. Probing the interferometer with coherent light, we systematically analyse the influence of experimental imperfections that could lead to a false-positive test result. In particular, we analyse the effects of detector nonlinearity, input-power and phase fluctuations on different timescales, closed-state transmissivity of shutters and crosstalk between different interferometer paths. In our experiment, a seemingly small shutter transmissivity in the order of about \num{2E-4} is the main source of systematic error, which suggests that this is a key imperfection to monitor and mitigate in future experiments.
\end{abstract}
\maketitle
%

\section{Introduction}
Quantum mechanics, like any theory in physics, is built upon a certain set of axioms. Even though they are by design not provable, they can be checked for consistency with nature by experiments. If the outcome does not falsify the axiom, it is most desirable to compare the experimental results with quantitative predictions of alternative theories, such that these alternatives can be ruled out or their parameter space restricted. For example, the Bell inequality \cite{Bell1964,Clauser1969} bounds correlations between pairs of entangled particles based on the assumption of local realism (i.e., the presence of hidden variables governing these correlations). Experimental violations of this bound \cite{Aspect1982}, especially when found in absence of loopholes \cite{Hensen2015,Giustina2015,Shalm2015,Rauch2018,Li2018} or with very high statistical significance \cite{Poh2015,Meraner2021}, allow ruling out hidden-variable models of quantum mechanics with high confidence.

Even on the more fundamental level of single-particle quantum mechanics, there is ongoing research probing the connection between wave functions and the probability as their measurable, physical quantity, as postulated by the axiom of Born's rule \cite{Born1926}. These experiments test for the existence of higher-order interferences \cite{Sorkin1994,Sinha2010,Hickmann2011,Park2012,Jin2017,Kauten2017,Cotter2017,Barnea2018,Pleinert2020}, which would contradict Born's rule or linearity and would require adaptations of quantum mechanics and/or alternative theories.

An axiom, which only very few experiments were able to probe so far, is the representation of quantum mechanical states by rays in a Hilbert Space based on complex numbers. In theory, states could be described using Hilbert spaces based on other fields, such as real, quaternion, octonion and possibly other hyper-complex numbers. Two notable theories using a different number system are real quantum mechanics \cite{Birkhoff1936,Stueckelberg1960} and \ac{QQM} \cite{Adler1995}. Both theories are mathematically consistent, but only quaternionic and (standard) \ac{CQM} are able to correctly describe the dynamics of a two-level system like the polarization state of a photon. In order to determine whether \ac{CQM} is sufficient or whether \ac{QQM} would better describe nature, specialised experiments are needed.

In 1979, Peres proposed a test to distinguish between quantum theories based on different number systems \cite{Peres1979}. This test is based on the dimension of the phase. While complex numbers can be represented by a single real phase parameter, three phase components are needed for quaternions. This allows distinguishing between complex and hypercomplex quantum mechanics in a scattering experiment with variable presence of one to three independent scatterers. In the same publication, he also proposed a specialized version based on the commutation of phase shifts. Only in the case of \ac{CQM} do such shifts commute. This version of the test has so far been implemented via the interference of neutrons \cite{Kaiser1984} and of single photons \cite{Procopio2017}, both showing no trace of quaternions whatsoever.
The proposed scattering experiment can also be recast as an interference scenario with three individually switchable amplitudes. Due to its intrinsic sensitivity to a variety of experimental imperfections, such an interferometric test requires a well-characterised high-precision setup and has so far not been implemented in any system.

In this work, we designed and set up an integrated optical waveguide interferometer, to be used in a Peres three-path interference test with photons. In order to access the required amplitude combinations, each path can be individually blocked by mechanical shutters. Compared to free-space setups \cite{Kauten2017} or partially integrated solutions \cite{Keil2016}, the present setup provides much better phase stability, which is a prerequisite to obtain meaningful results. For an in-depth analysis of potential systematic errors, which would bias a Peres test with single photons, we probed the interferometer with bright coherent light with negligible shot noise. While this will not let us distinguish between \ac{CQM} and \ac{QQM}, it enables identification of the dominant error mechanisms to be addressed before a Peres test with single photons can be run.

\section{Concept of the Peres test} 
In the standard formulation of quantum mechanics, a three-path interferometer with the paths A, B and C is fully described by three complex amplitudes with real phases $0\leq\phi_i\leq 2\pi$ (with $i \in \{\mathrm{A},\mathrm{B},\mathrm{C}\}$). Assuming full coherence, their respective phase differences $\Delta\phi_{ij} = \phi_i - \phi_j$ ($i\neq j$) must add to zero:
\begin{equation}
    \label{eqn:PhaseSum}
    \Delta\phi_\mathrm{AB} + \Delta\phi_\mathrm{BC} + \Delta\phi_\mathrm{CA} = 0.
\end{equation}
Applying the cosine function to \cref{eqn:PhaseSum} and using trigonometrical transformations results in the so-called Peres parameter
\begin{equation}
    \label{eqn:PeresParameter}
    F \defeq \alpha^2 + \beta^2 + \gamma^2 - 2 \alpha \beta \gamma = 1,
\end{equation}
with $\alpha \defeq \cos(\Delta\phi_\mathrm{BC})$, $\beta \defeq \cos(\Delta\phi_\mathrm{CA})$ and $\gamma \defeq \cos(\Delta\phi_\mathrm{AB})$ describing the normalized interference terms. 
The description of quantum mechanical waves with hypercomplex numbers increases the dimension of their phase. In the case of quaternions the phase becomes three-dimensional, such that \cref{eqn:PeresParameter} does not necessarily hold, which is the starting point for the formulation of the interferometric Peres test. Deviations from the expected value $F=1$ are, according to \cite{Peres1979}, interpreted as
\begin{equation}
    F=\cases{<1 & hypercomplex numbers are admissible\\
    >1 & superposition principle is violated\\}.
\end{equation}
In order to connect this parameter with directly measurable probabilities, the photon detection rates $P_{\mathrm{A}}, P_{\mathrm{B}}, P_{\mathrm{AB}}$ are measured at one of the interferometer's outputs for various path transmission settings (A open, B open, A and B open, etc.) and constant input flux. In the case of a two-path interferometer with paths A and B, the measured photon rate $P_\mathrm{AB}$ is given by
\begin{equation}
    P_\mathrm{AB} = P_\mathrm{A} + P_\mathrm{B} + 2 \sqrt{P_\mathrm{A} P_\mathrm{B}} \gamma
\end{equation}
This equation can be used to express the normalized interference term as
\begin{equation}
    \label{eqn:gamma}
    \gamma = 
    \frac{P_\mathrm{AB} - P_\mathrm{A} - P_\mathrm{B}}{2 \sqrt{P_\mathrm{A} P_\mathrm{B}}}. 
\end{equation}
The equations for $\alpha$ and $\beta$ can be derived analogously and give 
\begin{eqnarray} 
    \label{eqn:alpha}
    \alpha = 
    \frac{P_\mathrm{BC} - P_\mathrm{B} - P_\mathrm{C}}{2 \sqrt{P_\mathrm{B} P_\mathrm{C}}} \hspace{1em}\text{and} \\
    \label{eqn:beta}
    \beta = 
    \frac{P_\mathrm{CA} - P_\mathrm{A} - P_\mathrm{C}}{2 \sqrt{P_\mathrm{A} P_\mathrm{C}}}. 
\end{eqnarray}
In order to obtain all normalized interference terms and calculate $F$, all shutter combinations with a single open path or two open paths need to be measured.

Various experimental imperfections like input-power fluctuations, phase fluctuations and non-zero transmissivity of the shutter can lead to deviations from $F=1$. For a meaningful interpretation of the Peres experiment, it is therefore necessary to characterise or measure all these imperfections of the interferometer as precisely as possible (see \cref{sec:SysErrors}).

\section{Experimental setup}
\label{sec:Setup}
\begin{figure}[htb]
	\includegraphics[scale=0.45]{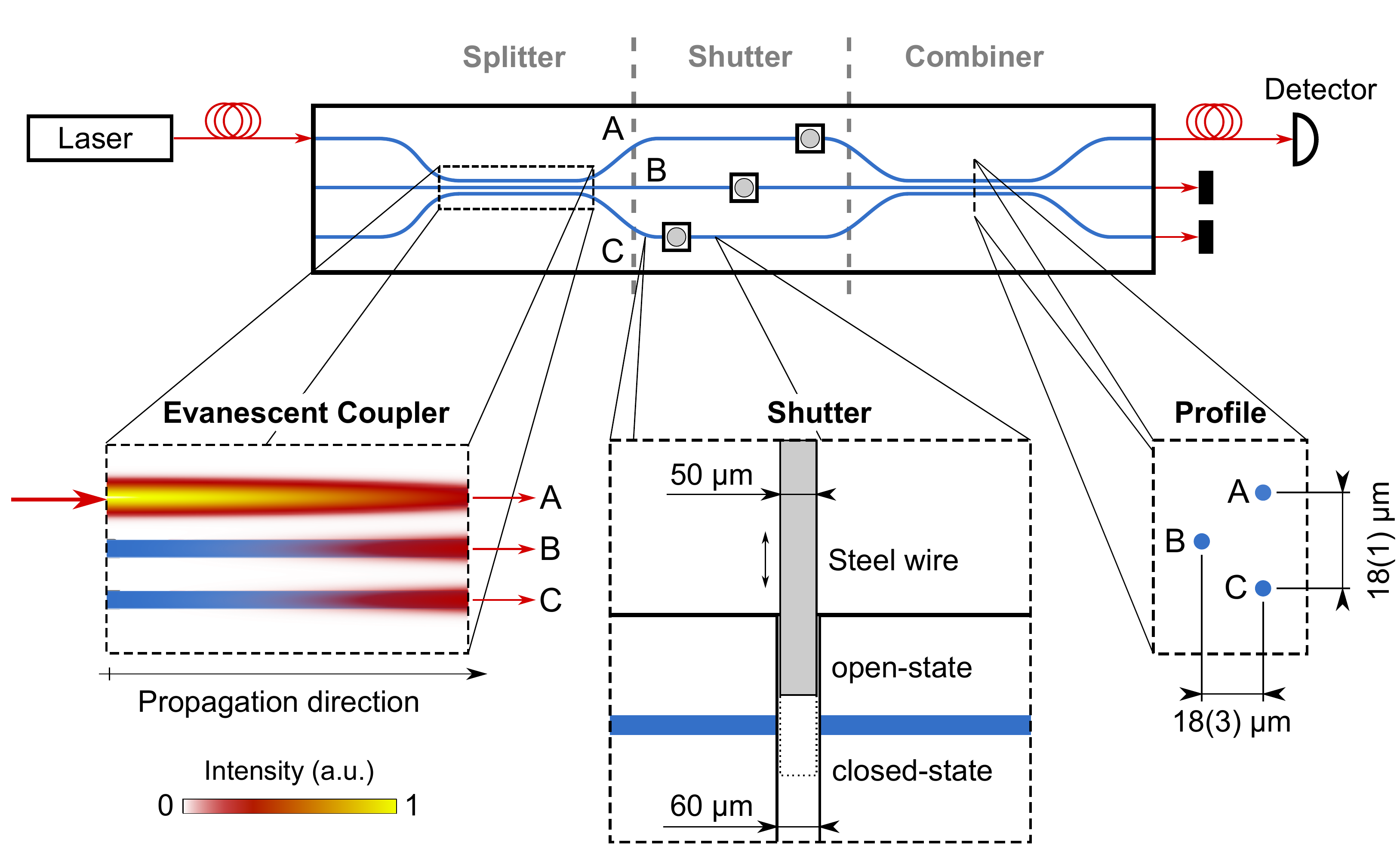}
	\caption{Schematics of the three-path Mach-Zehnder interferometer in a fused silica chip with mechanical shutters. A \SI{808}{\nano\meter} cw laser is coupled via a fiber into one of the three waveguide paths A, B and C (shown in blue) and split equally among them via evanescent coupling (left inset). Holes through each waveguide path allow steel wires to act as shutters (middle inset). The light in each path is then recombined using evanescent coupling. The three paths A, B and C are aligned as a triangle (right inset).}
	\label{fig:Setup}
\end{figure}

To perform the Peres test we used a three-path Mach-Zehnder interferometer integrated in a fused silica glass chip, with external mechanical shutters, as can be seen in \cref{fig:Setup}. The waveguides were directly inscribed using femtosecond laser pulses, which enables three-dimensional path designs \cite{Szameit2010,Meany2015}. The three waveguides were aligned in a triangle, which was continuously rescaled in a form-preserving manner to produce the various functional elements of the chip. At minimal size (base length=\SI{18(1)}{\micro\meter}), the reduced spacing between each waveguide leads to evanescent coupling (see the left inset of \cref{fig:Setup}). 
The length of the evanescent coupler was chosen to yield for each input an approximately balanced output splitting between the three ports \cite{Spagnolo2013}.
One such unit opens the interferometer (henceforth termed splitter), while a mirror-inverted copy (combiner) closes it. In the region between splitter and combiner, the triangle has a maximal baselength of \SI{201(1)}{\micro\meter}, minimizing cross-coupling between the separate waveguides. These straight paths are intersected by rectangular \SI{60x60}{\micro\meter} holes, produced by laser-inscription of nanocracks (in the same fabrication step as the waveguides) and subsequent selective chemical etching with HF-acid \cite{Marcinkevicius2001,Hnatovsky2006,Crespi2012}. 
Steel wires with \SI{50}{\micro\meter} diameter were inserted into these holes to act as mechanical shutters, whose open/closed state depends on the vertical position of the wire. In order to reduce losses from refraction and reflection for the open shutter position, the holes were filled with an index matching oil, with a refractive index difference of $\Delta n \approx \num{-5E-3}$ at \SI{23}{\degreeCelsius} in comparison with the fused silica.

The in- and output paths of the chip were connected to optical fibers via fiber-arrays with refractive index matching gel bridging the gap between the fiber end and the chip surface. The gel 
has a refractive index difference of $\Delta n \approx \num{-3E-3}$ with respect to the fused silica. On the input we used a polarization-maintaining fiber and on the output side a single-mode fiber, in order to reduce the detected stray-light in comparison to using a multimode or free-space imaging.
The light of the output fiber is collimated onto a photodiode (Physimetron A139-001). As light source we used an \SI{808}{\nano\meter} laser diode (Thorlabs LD808P030), which is temperature, power stabilised. The power stabilisation is performed by a liquid crystal noise eater (Thorlabs LCC3112), which results in a relative power stability (standard deviation) of \SI{0.32}{\percent} over \SI{9.2}{\hour}.

The close proximity of the waveguides in the evanescent couplers can lead to a rotation of the polarization, which depends on the orientation of the waveguides in the triangle relative to the orientation of the elliptical waveguide profiles \cite{Heilmann2014}. As the detector is polarization insensitive and not every path experiences the same rotation, a polarizer was placed in front of the detector to project the output light into a single polarization and, thereby, prevent systematic errors in our measurements (see \cref{sec:polarization}).

The whole fused silica chip was enclosed in an aluminium housing, whose bottom and sides are insulated with polystyrene sheets, with gaps for the shutter holes and the in- and output ports of the photonic circuit. The housing was temperature stabilized so that the largest observable fluctuations were smaller than \SI{E-3}{\kelvin} by a proportional–integral-controlled resistive heater at the bottom of the housing, centered below the shutter holes. The temperature sensor was mounted on the outside of the housing and was almost completely covered by the insulation. As the top part of the housing was not insulated, the heater created a vertical temperature gradient. Due to the triangular arrangement of the three interferometer paths, this led to a relative phase change between path B (tip of the triangle) and paths A and C (base of the triangle). Therefore, it was possible to slightly tune the phase differences $\Delta\phi_\mathrm{AB}$ and $\Delta\phi_\mathrm{BC}$ simultaneously by changing the housing temperature. They changed by roughly \SI{-2.5E-2}{\radian\per\kelvin} in the range from \SIrange{25}{35}{\degreeCelsius}.

The housing was mounted with nylon screws onto machinable glass-ceramic posts, to reduce the heat flow at the bottom of the housing and, thereby, increase the strength of the temperature gradient. Furthermore, these posts reduced the influence of temperature changes on the in- and output coupling efficiency of the chip to the fiber arrays, thanks to their lower thermal expansion coefficient in comparison to stainless steel posts. We inserted indium foil between the chip and its housing to eliminate air gaps and increase the thermal conductance.
%

\section{Experimental results}
\label{sec:expResults}
We measured the transmitted optical power for all eight shutter combinations in random order and repeated these cycles multiple times (each time with a new independent random ordering to average over possible long-term drifts). The configuration with all three shutters closed $P_0$ represents the experimental background, e.g. stray-light or detector dark-current. We subtracted the measured background from the other combinations of the same cycle to correct for these influences. The shutter configuration of all paths open $P_\mathrm{ABC}$ is only needed to analyse specific systematic errors as detailed in \cref{sec:crosstalk}. We performed the Peres test for two different temperature settings of the chip housing, \SI{23}{\degreeCelsius} and \SI{30}{\degreeCelsius}, respectively. The temperature stability is \SIrange{3}{9}{\milli\kelvin} (standard deviation) for timescales in the order of the presented measurements. The two different temperatures correspond to different phase differences between the waveguides at the base and the one at the tip of the triangular path arrangement. The results of both measurements were filtered for shutter malfunctions. The durations of the two measurements were \SI{9.2}{\hour} and \SI{13.2}{\hour} for \SI{23}{\degreeCelsius} and \SI{30}{\degreeCelsius}, respectively.

\begin{figure}[htb]
	\begin{tikzpicture}
		\node (img1) {\includegraphics[scale=0.605]{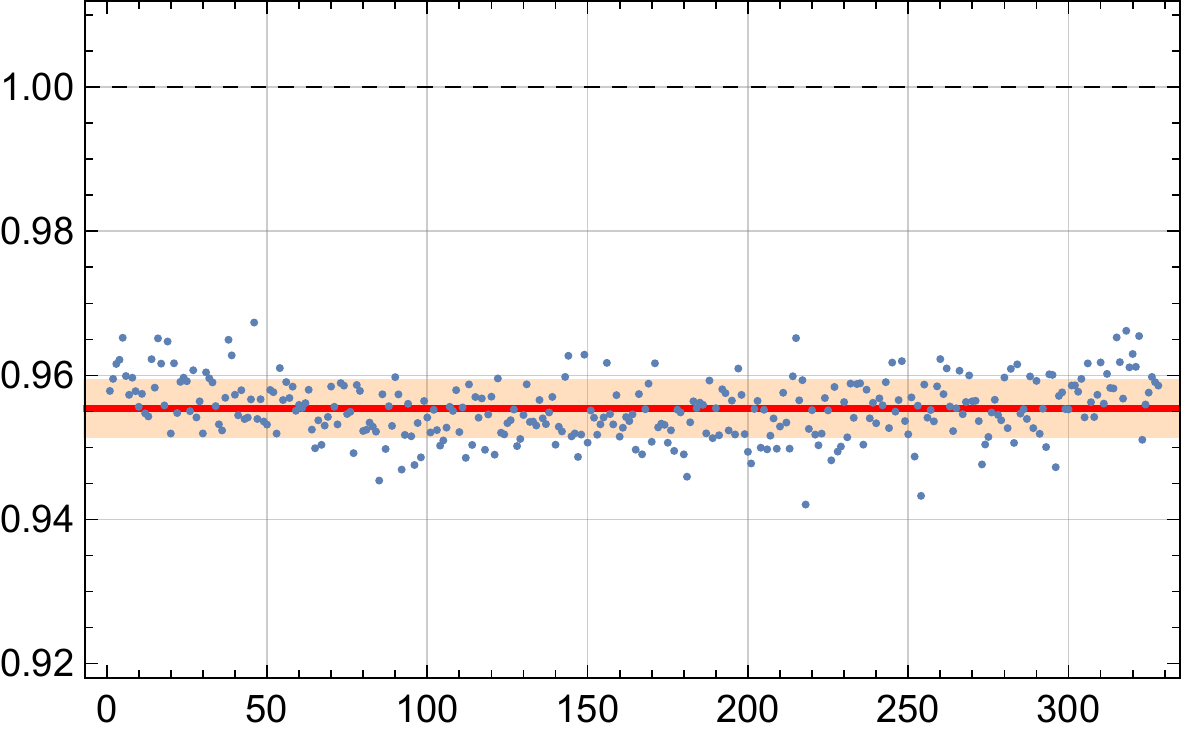}};
		\node[right	=0 of img1]	(img2) {\includegraphics[scale=0.605]{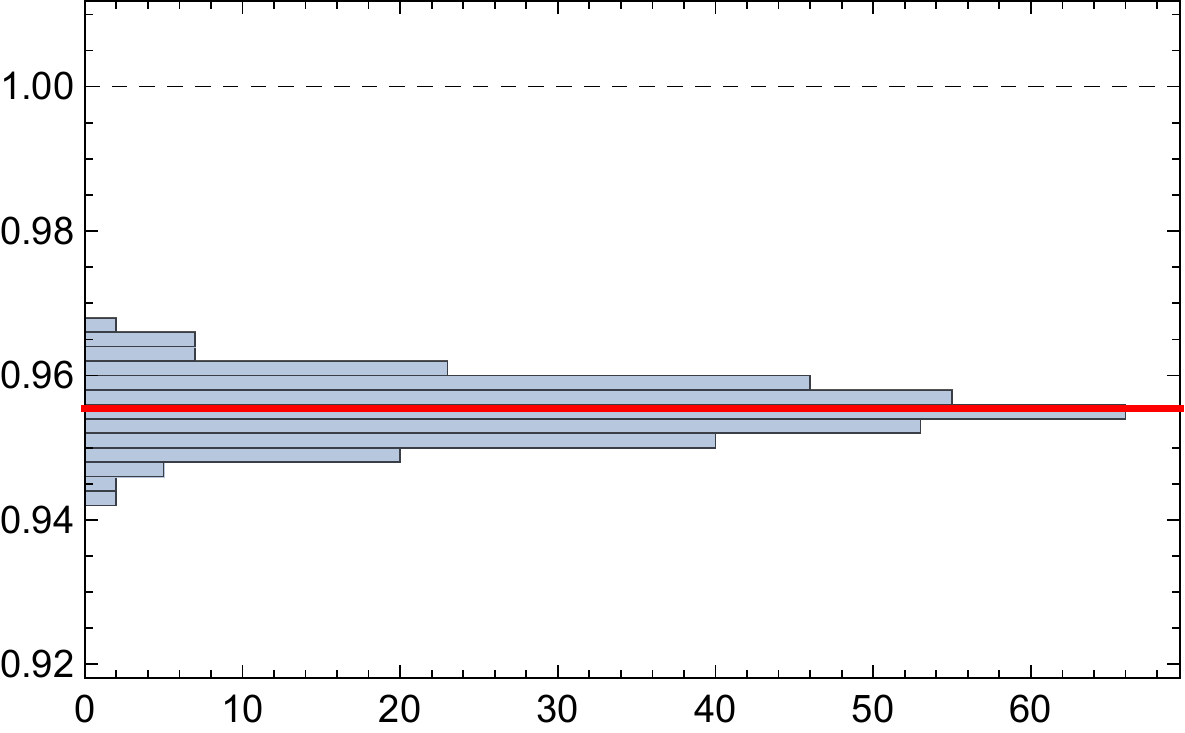}};
		\node[below =0 of img1] (img3) {\includegraphics[scale=0.605]{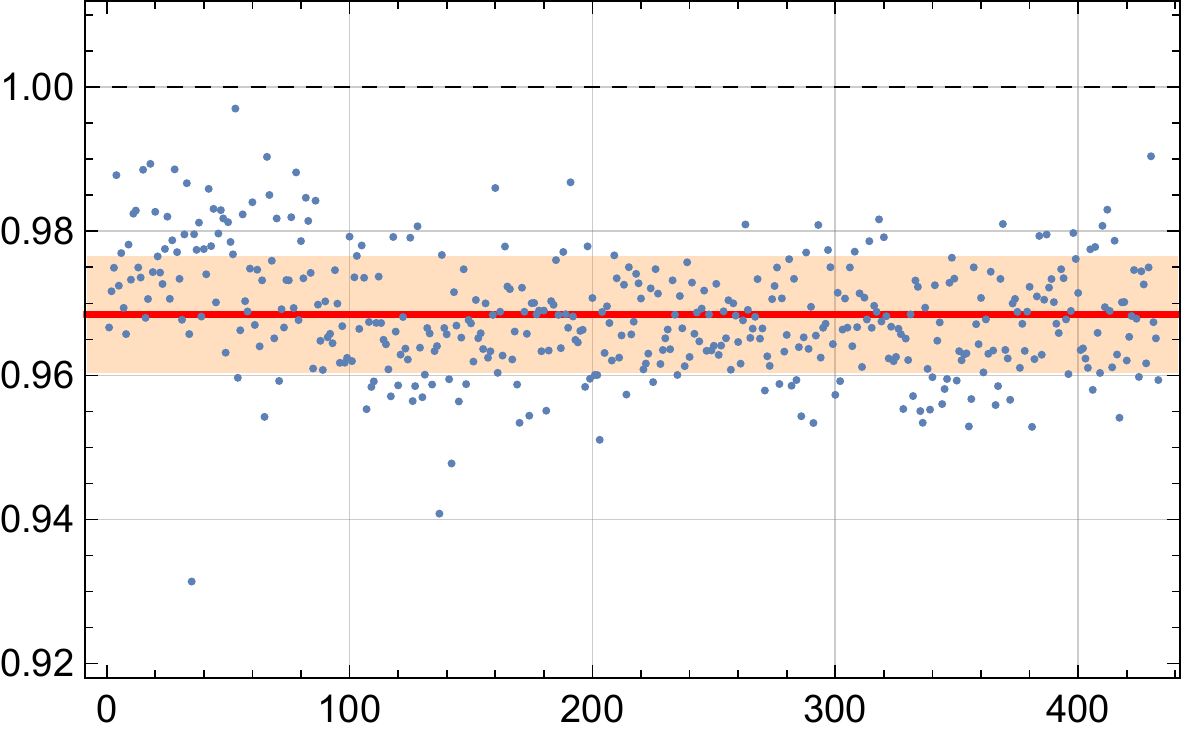}};
		\node[right	=0 of img3]	(img4) {\includegraphics[scale=0.605]{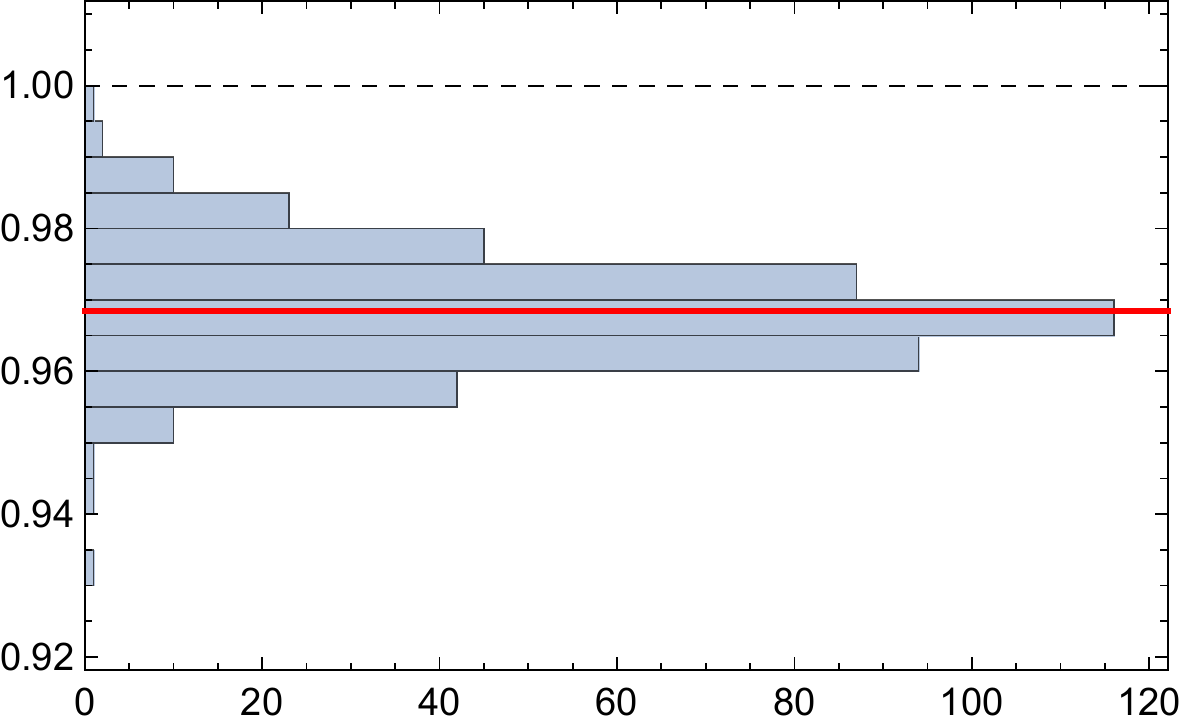}};
		\node[left=0 of img1, rotate=90, anchor=center,yshift=0.2cm] {$F$};
		\node[left=0 of img3, rotate=90, anchor=center,yshift=0.2cm] {$F$};
		\node[below=0 of img3, anchor=center, yshift=-0.2cm, xshift=0.4cm]	{$k$};
		\node[below=0 of img4, anchor=center, yshift=-0.2cm, xshift=0.4cm]	{Frequency};
		\node[above right=-0.75 of img1, anchor=north east, yshift=0.2cm, fill=white] {\SI{23}{\degreeCelsius}}; 
		\node[above right=-0.75 of img2, anchor=north east, yshift=0.2cm, fill=white] {\SI{23}{\degreeCelsius}};
		\node[above right=-0.75 of img3, anchor=north east, yshift=0.2cm, fill=white] {\SI{30}{\degreeCelsius}};
		\node[above right=-0.75 of img4, anchor=north east, yshift=0.2cm, fill=white] {\SI{30}{\degreeCelsius}};
	\end{tikzpicture}
	\caption{Peres parameter $F$ (in blue) for multiple successive measurements, with index $k$, on the left and the corresponding histogram on the right. The plots of each row were measured at different housing temperature as stated in the top right of each plot. The red horizontal lines describe the mean values and the orange shaded bands represents one standard deviation. The expected value of $F=1$ is indicated by a black dashed line.
	}
	\label{fig:peres_param}
\end{figure}
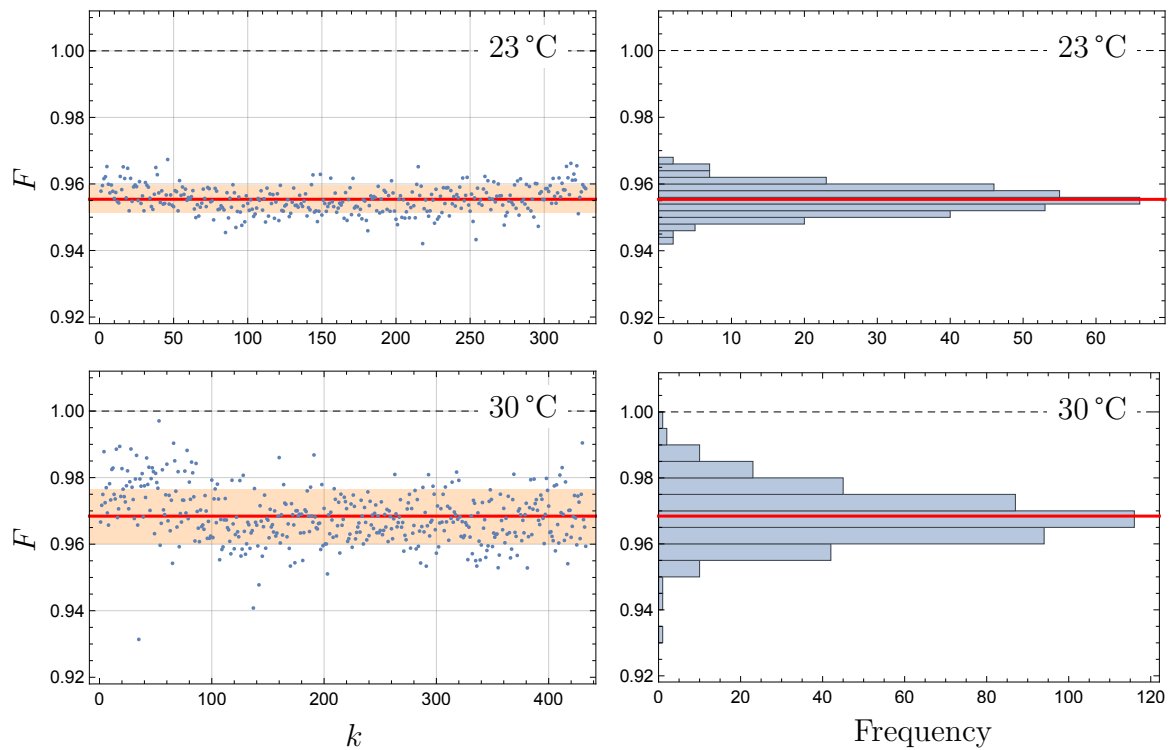

\Cref{fig:peres_param} shows the result of the Peres test for each cycle at the two temperature settings of \SI{23}{\degreeCelsius} and \SI{30}{\degreeCelsius}, respectively. The mean of the measured Peres parameter $F$ is
\begin{equation}
    \mean{F_{\SI{23}{\degreeCelsius}}} = \num{0.9553(4)} \hspace{0.5em} \text{and} \hspace{0.5em} \mean{F_{\SI{30}{\degreeCelsius}}} = \num{0.9683(9)},
\end{equation}
where the indicated uncertainties are the standard errors of the mean corrected for the autocorrelation according to \cite{Zhang2006}. Both results show a significant deviation in the order of \num{-E-2} from the expected value of $F=1$, which we attribute to systematic errors of the experiment. We will go into more detail about possible sources of such errors in \cref{sec:SysErrors}.
%

\section{Systematic error analysis}
\label{sec:SysErrors}
The Peres parameter is influenced by various experimental imperfections, e.g. fluctuations of the input power and cross-influence of shutters. How much any of these effects influence the result depends on the prevailing phase differences of the interferometer paths. These can be directly reconstructed from the measured combinations using \cref{eqn:gamma}-\eref{eqn:beta}. However, due to the very same error sources influencing the Peres test, the total of the so-obtained phase differences is not zero. In order to estimate the original uninfluenced set of phases for each temperature setting, we calculated the closest set of phases, that is consistent with \cref{eqn:PhaseSum} (see \cref{sec:phaseSpaceReconstr} for details).
With this set of corrected phases we can then apply various models of experimental imperfections and determine their influence on the measured Peres parameter. The directly reconstructed and the corrected phase differences are listed as normalised interferences terms (cosine of the phase difference) in \cref{tab:phaseSpaceReconstr}. 
\begin{table}
	\caption{\label{tab:phaseSpaceReconstr}Directly reconstructed and corrected sets of normalised interference terms $\{\alpha, \beta, \gamma\}$ averaged over all cycles for the two housing temperatures $T_\mathrm{h}$. The corrected sets satisfy \cref{eqn:PhaseSum} with minimal Euclidean distance to the directly reconstructed sets. All given errors are the standard deviation over all cycles.}
	\begin{indented}
    \lineup
    \item[]
	\begin{tabular}{crccc}
	\toprule
    $T_\mathrm{h}$ &   & $\mean{\alpha}$    & $\mean{\beta}$ & $\mean{\gamma}$ \\
	\midrule
	\SI{23}{\degreeCelsius}    & \twoline{l}{reconstructed}{corrected}    & \twoline{l}{\num{-0.765(16)}}{\num{-0.806(15)}} & \twoline{l}{\num{ 0.941(5) }}{\num{ 0.961(3)}} & \twoline{l}{\num{-0.664(15)}}{\num{-0.612(15)}}  \\ \addlinespace
	\SI{30}{\degreeCelsius}    & \twoline{l}{reconstructed}{corrected}    & \twoline{l}{\num{-0.405(19)}}{\num{-0.449(18)}} & \twoline{l}{\num{ 0.980(5) }}{\num{ 0.989(3)}} & \twoline{l}{\num{-0.355(16)}}{\num{-0.310(20)}} \\
	\bottomrule
	\end{tabular}
    \end{indented}
\end{table}
In the following, we will analyse a number of effects that lead to a systematic shift of the Peres parameter. In general these effects will also lead to random errors.

\subsection{Detector nonlinearity}
\label{sec:DetectorNonLinearity}
In tests searching for higher-order interferences, any nonlinearity in the experiment can lead to a false-positive detection of higher-order interferences \cite{Kauten2017,Rozema2020}. Similarly, nonlinearities also have a systematic effect on the Peres test, because they bias the addition of the rates in \cref{eqn:gamma}-\eref{eqn:beta}. In our experiment, which uses low-power continuous-wave illumination, the photodetector and its subsequent electronics are the dominant sources of nonlinearity, which need to be accounted for. Therefore, we measured the nonlinearity of the detector in independent beam combination experiments, following the procedure outlined in \cite{Kauten2014}. Correcting the measured data of the Peres experiment for this detector nonlinearity results in a correction of $F$ by
\begin{equation}
    \Delta F\scr{NL}_{\SI{23}{\degreeCelsius}} = \num{-4.7E-5} \hspace{0.5em} \text{and} \hspace{0.5em}
    \Delta F\scr{NL}_{\SI{30}{\degreeCelsius}} = \num{-9.4E-5}.
\end{equation}
These deviations are negligible compared to the observed deviation. Thus, we can rule out nonlinearity as a relevant source of error in our experiment.

\subsection{Fluctuations}
\label{sec:fluctuations}
The measurements shown in this work were taken over periods of \SIlist{9;13}{\hour} and temporal effects can occur on various timescales. The effect of slow drift in power or phase difference on a timescale much longer than a measurement cycle ($\sim \SI{1.75}{\minute})$ is reduced by measuring all eight path combinations in random order. Power fluctuations occurring on timescales shorter than the measurement duration for one shutter combination ($\sim \SI{13}{\second}$) are averaged over multiple successive data points. Phase fluctuations on that timescale, however, reduce the interference contrast, which is considered in \cref{sec:InterfContrast}. In this section, we concentrate on power and phase fluctuations on the intermediate timescale of one measurement cycle. 

In a first scenario, we assumed fluctuations of the input power during a cycle and overall perfect phase stability. To simulate this, we expressed the power $P_i$ of each single path combination $i \in \{\mathrm{A},\mathrm{B},\mathrm{C}\}$ as
\begin{equation}
    \label{eqn:PowFluc_onePathTerm}
    P_i = P_{\mathrm{in},i} \cdot T_i.
\end{equation}
The variable $P_{\mathrm{in},i}$ is the average power coupled into the chip during the measurement of the combination $i$ and $T_i$ the transmission along this path to the detector. For the two-path combinations $i,j \in \{\mathrm{A},\mathrm{B},\mathrm{C}\}$, with $i \neq j$, the power $P_{ij}$ is given by 
\begin{equation}
    \label{eqn:PowFluc_twoPathTerm}
    P_{ij} = P_{\mathrm{in},ij} \left(T_i + T_j + 2 \sqrt{T_i T_j} \cos(\Delta\phi_{ij})\right).
\end{equation}
The transmissions were measured in a separate experiment as $T_\mathrm{A} = 0.26$, $T_\mathrm{B} = 0.52$ and $T_\mathrm{C} = 0.22$, while the phase differences were calculated from the corrected normalised interference terms (see \cref{tab:phaseSpaceReconstr}). Each input power is treated as an independent normally distributed random variable with mean and standard deviation extracted from the measured data (for details see \cref{sec:CalcPhaseFluc}). Based on the measurement data we can determine the standard deviation of the fluctuations on either the timescale of a single shutter setting or the timescale of the whole measurement, but not directly on the scale of one cycle. 
The standard deviation on the timescale of the whole measurement is larger and thus leads to more deviations in the Peres parameter. We therefore use it as an upper-bound estimation for the effect of input-power fluctuations. We calculated the deviations of the Peres Parameter from the expected value $\Delta F = F-1$ and its spread $\sigma_F$ with a Monte Carlo simulation.
\begin{table}
	\caption{\label{tab:PowANDPhaseFluc}Influence on the Peres Parameter $F$ caused by power and phase fluctuations, respectively. The mean of the distribution calculated with a Monte Carlo simulation is given as a deviation from the expected value $\Delta F = F - 1$. The standard deviation is described by $\sigma_F$. The results are listed for the two datasets at different housing temperatures $T_\mathrm{h}$.}
	\begin{indented}
    \lineup
    \item[]
	\begin{tabular}{lrrcrr}
	\toprule
	& \multicolumn{2}{c}{Power Fluctuations}	&& \multicolumn{2}{c}{Phase Fluctuations}  \\ \cmidrule{2-3}\cmidrule{5-6}
	\tcellc{$T_\mathrm{h}$}  & \tcellc{$\Delta F\scr{Pow}$}    & \tcellc{$\sigma_{F\scr{Pow}}$} && \tcellc{$\Delta F\scr{Phase}$}    & \tcellc{$\sigma_{F\scr{Phase}}$} \\
	\midrule
	\SI{23}{\degreeCelsius} & \num{5.1E-4}   & \num{1.8E-2} && \num{ 0.8E-4}   & \num{1.0E-2}\\
	\SI{30}{\degreeCelsius} & \num{1.4E-4}   & \num{1.4E-2} && \num{-8.1E-4}   & \num{1.4E-2}\\
	\bottomrule
	\end{tabular}
    \end{indented}
\end{table}
The results in \cref{tab:PowANDPhaseFluc} show that power fluctuations mainly lead to a random uncertainty of the measured Peres parameter and only a small systematic shift.

To estimate the impact of phase fluctuations, we considered fluctuations on a timescale slower than one shutter combination. Any faster fluctuations would lead to a reduction of the interference contrast, which we discuss in \cref{sec:InterfContrast}. 
To model the slower phase fluctuations we used \cref{eqn:PowFluc_onePathTerm,eqn:PowFluc_twoPathTerm} and assumed perfect stability of the input power for each combination $P_{\mathrm{in},ij} = P_{\mathrm{in},i} = P_\mathrm{in}$. The phase differences of each two-path combination are modelled by independent Gaussian distributed numbers with mean values given by the corrected phase differences (see \cref{tab:phaseSpaceReconstr}) and the standard deviation given by the observed fluctuations on the timescale of the whole measurement (see \cref{sec:CalcPhaseFluc} for details).
We again performed a Monte Carlo simulation and obtained the values for $\Delta F$ and $\sigma_F$ as given in \cref{tab:PowANDPhaseFluc}. As it turns out, the phase fluctuations have a similar impact as the power fluctuations, i.e. they mainly lead to a larger random uncertainty and only to a small systematic shift of the Peres parameter. The latter is for both types of fluctuations two orders of magnitude smaller than the observed $\Delta F$ and therefore negligible.

\subsection{Interference contrast}
\label{sec:InterfContrast}
Phase fluctuations that are faster than the duration of one shutter setting lead to a reduction of the normalised interference terms $\chi \in \{\alpha, \beta, \gamma\}$. Our model of this effect describes the measured normalise interference terms $\tilde \chi$ as 
\begin{equation}
    \chi \hspace{2em} \longrightarrow \hspace{2em} \tilde \chi = C_\chi \cdot \chi,
\end{equation}
where $C_\chi<1$ is the reduced interference contrast.
This model predicts a strong dependence of the Peres parameter on the interference contrast $C_\chi$ of each pair of paths. Assuming the same interference contrast for all three interference terms ($C_\chi = 1- \Delta C$), the deviation of the Peres parameter is calculated as follows:
\begin{equation}
    \Delta F\scr{C} = 2 \left( \alpha \beta \gamma - 1 \right) \Delta C - \left( 4 \alpha \beta \gamma - 1 \right) \Delta C^2 + 2 \alpha \beta \gamma \cdot \Delta C^3
\end{equation}
To estimate the reduction of the interference contrast, we used the corrected phase fluctuations on the timescale between a shutter setting ($\sim \SI{13}{\second}$) and a single measurement ($\sim \SI{0.11}{\second}$). How these were extracted from the measured data is outlined in \cref{sec:CalcPhaseFluc}.
We used these fluctuations to calculate the interference contrasts with a Monte Carlo simulation around the mean phase difference $\Delta\phi=0$ for each interference term. Due to the maximum of the cosine function at this point, phase fluctuations lead to a unidirectional reduction of the power, such that the resulting interference contrast serves as an estimate at all possible mean phases.
However this method does not consider the effect of fluctuations on a timescale below a single measurement (above $\sim \SI{9}{\hertz}$). As worst a case estimate, we used the smallest of these interference contrasts ($C=\num{0.999998}$) for each normalised interference term $\tilde\chi$ resulting in a deviation from the Peres parameter $\Delta F^\mathrm{C}$ in the order $\num{E-6}$ for both measurements. Compared to the measured deviations this effect is negligible.

We also separately measured the interference contrast of our setup for one path combination by thermally tuning the phase difference, which gives a contrast of $C_\alpha = \num{1.000(29)}$. The measurement and its evaluation is described in \cref{sec:InterfContrMeasurement}. Note that the number of subsequent measurements per shutter setting is only $1/10$ of the number in the main experiment. This leads to reduced duration of each shutter setting and therefore reduces the range of fluctuation frequencies affecting the contrast. Hence this interference contrast probably underestimates the degradation of contrast from fluctuations.

\subsection{Crosstalk}
\label{sec:crosstalk}
Ideally, the state of each shutter should only influence the path it is inserted to. However, experimentally one cannot exclude the possibility that closing one shutter also affects the light in the other two paths. The moving steel wires could for example lead to bending of the chip and with it to a change of the phase of the other paths. In theory, the shutter could also influence the transmitted power of other paths, however this seems less likely than an influence on the phase and is, therefore, not considered here.

\begin{figure}[htb]
	\centering
	\def\svgwidth{0.40\textwidth}
	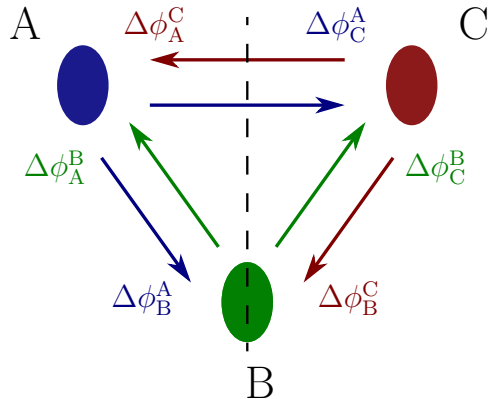
	\caption{Three slightly elliptical waveguides A, B and C in a triangular arrangement, with the corresponding phase crosstalk parameters $\Delta\phi^i_j$, where $i$ is the shutter affecting path $j$. The dashed line represents the symmetry axis.}
	\label{fig:PhaseCrosstalk}
\end{figure}

In the following model of phase crosstalk, the closed shutter of path $i$ changes the phase of another path $j$ by $\Delta\phi^i_j$ as opposed to the open state in path $i$. This is shown for our triangular path arrangement in \cref{fig:PhaseCrosstalk}.
We constrained these six parameters by symmetry considerations: Due to the slightly elliptical waveguide profiles, which are common for femtosecond laser writing, the symmetry axis of the interferometer runs vertically through path B (cf. \cref{fig:PhaseCrosstalk}). Therefore, it seems very plausible to impose this mirror symmetry also on the crosstalk terms. That is, we assumed that the crosstalk originating from path A is equal to the one from path C and the crosstalk originating from B is the same for both of the other paths:
\begin{equation}
    \eqalign{
    \Delta\phi^\mathrm{A}_\mathrm{C} = \Delta\phi^\mathrm{C}_\mathrm{A} \defeq \Delta\phi_\mathrm{H},\\
    \Delta\phi^\mathrm{A}_\mathrm{B} = \Delta\phi^\mathrm{C}_\mathrm{B} \defeq \Delta\phi^\mathrm{CA}_\mathrm{D},\\
    \Delta\phi^\mathrm{B}_\mathrm{A} = \Delta\phi^\mathrm{B}_\mathrm{C} \defeq \Delta\phi^\mathrm{B}_\mathrm{D}.}
\end{equation}
We further assume the two diagonal phase crosstalks $\Delta\phi^\mathrm{B}_\mathrm{D}$ and $\Delta\phi^\mathrm{CA}_\mathrm{D}$ to be of same magnitude
\begin{equation}
    \Delta\phi^\mathrm{B}_\mathrm{D} = \Delta\phi^\mathrm{CA}_\mathrm{D} \defeq \Delta\phi_\mathrm{D}.
\end{equation}
These simplifications lead to the following equations for the normalized interference terms
\begin{equation}
    \label{eqn:crosstalkNormInterf}
    \eqalign{
    \alpha  = \cos(\Delta\phi_\mathrm{BC} + \Delta\phi_\mathrm{DH}),\\
    \beta   = \cos(\Delta\phi_\mathrm{CA}),\\
    \gamma  = \cos(\Delta\phi_\mathrm{AB} - \Delta\phi_\mathrm{DH}),}
\end{equation}
with $\Delta\phi_\mathrm{DH} = \Delta\phi_\mathrm{D} - \Delta\phi_\mathrm{H}$. The effects on the normalized interference term $\beta$ cancel each other out, leaving only the parameter $\Delta\phi_\mathrm{DH}$ to be quantified. 
For this purpose, we additionally performed the Sorkin test, which probes for higher-order interferences in quantum mechanics (see \cite{Sorkin1994,Sinha2010,Hickmann2011,Park2012,Jin2017,Kauten2017,Cotter2017,Barnea2018,Pleinert2020} and \ref{sec:Sorkin_Appendix}), on the same set of data used for the Peres test. The Sorkin test has the advantage that of all effects considered in this chapter only crosstalk and nonlinearity (which can here be excluded due to its miniscule effect on the measured data; cf. \cref{sec:DetectorNonLinearity}) can lead to a systematic shift in its result, $\epsilon$ \cite{Kauten2017}. Therefore, the deviation from the expected value $\epsilon = 0$ can be used to estimate the magnitude of the crosstalk present in our setup. The measured $\epsilon$ (see \cref{sec:Sorkin_Appendix}) deviate significantly from $\epsilon=0$ for both temperatures. Using our crosstalk model one can then calculate the crosstalk strength $\Delta\phi_\mathrm{DH}$ required to produce the observed deviation. Based on these results we then calculated the expected deviation of the Peres parameter $\Delta F$ as listed together with $\Delta\phi_\mathrm{DH}$ in \cref{tab:Crosstalk}.
\begin{table}
	\caption{\label{tab:Crosstalk}Calculated deviation of the Peres parameter from its expected value $\Delta F^\mathrm{CT}=F^\mathrm{CT}-1$ due to phase crosstalk with a phase shift of $\Delta\phi_\mathrm{DH}$. The results are given for the two measurements at different housing temperatures $T_\mathrm{h}$.}
	\begin{indented}
    \lineup
    \item[]
	\begin{tabular}{crr}
	\toprule
	 $T_\mathrm{h}$    & \tcellc{$\Delta\phi_\mathrm{DH}$ (\num{E-2} \si{\radian})}  & \tcellc{$\Delta F\scr{CT}$ (\num{E-3})}   \\
	\midrule
	\SI{23}{\degreeCelsius}  & \num{-1.7}    & \num{-9.0} \\
	\SI{30}{\degreeCelsius}  & \num{-1.4}    & \num{ 6.6} \\
	\bottomrule
	\end{tabular}
    \end{indented}
\end{table}
Note that we also tested an asymmetric model, where all phase crosstalk arises from a single path (e.g. from one wire moving under stronger friction than the others). This delivered consistently smaller $\abs{\Delta F}$. The results of this section indicate that some crosstalk is present in our experiment. It could well lead to a non-negligible bias on the Peres test, yet the size of this effect seems to fall short of the observed deviations $\abs{\Delta F}$ by one order of magnitude.

\subsection{Residual light}
Unfortunately, our wire-based shutters exhibit finite transmissivity in their closed state. Therefore we assign to the shutter in each path $i$ a closed-state transmissivity $\tau_{i}$, due to residual light passing through the wire-blocked hole. This additional light can interfere with the light from open paths and influence the measured Peres parameter. To simplify things, we consider the same transmissivity for each shutter ($\tau_i = \tau \ \forall i \in \{\mathrm{A},\mathrm{B},\mathrm{C}\}$) and the phase of the residual light to be shifted with respect to the phase in the open-state by $\phi_\mathrm{Sh}$. With this model the measured background $P_0$ (all shutters closed), the single path $P_i$ and the two path combinations $P_{ij}$ are described by 
\begin{eqnarray}
    \label{eqn:P0_resLight}
    P_0 =& \tilde P_0 + P_\mathrm{in} \tau \left( \sum_i T_i + \sum_{i,j} (1-\delta_{i,j})\sqrt{T_i T_j} \cos(\Delta\phi_{ij}) \right)\\
    P_i =& \tilde P_0 + P_\mathrm{in} \left(T_i + \tau \sum_{j \neq i} T_j + 2\sqrt{\tau T_i} \sum_{j \neq i} \sqrt{T_j} \cos(\Delta\phi_{ij}+\phi_\mathrm{Sh}) + \right. \\
         &+ \left. 2 \tau \sqrt{T_k T_l} \cos(\Delta\phi_{kl}) \right) \nonumber\\
    P_{ij} =& \tilde P_0 + P_\mathrm{in} \left(T_i + T_j + \tau T_k + 2 \sqrt{T_i T_j} \cos(\Delta\phi_{ij}) + \right.\\
         &+ \left.2 \sqrt{\tau T_i T_k} \cos(\Delta\phi_{ik} + \phi_\mathrm{Sh}) + 2 \sqrt{\tau T_j T_k} \cos(\Delta\phi_{jk} + \phi_\mathrm{Sh})\right), \nonumber
\end{eqnarray}
with $i,k,l \in \{\mathrm{A},\mathrm{B},\mathrm{C}\}$ and $i \neq k \neq l$. The input power $P_\mathrm{in}$ already includes in- and out-coupling efficiencies. The transmissivities $T_i$ are the percentage of light passing through path $i$ and reaching the detector. The power $\tilde P_0$ describes a generic shutter-independent incoherent background, e.g. from straylight or detector dark current. This term subsequently drops out, when calculating the normalised interference terms, as the measured background is subtracted from the single- and two-path combinations.

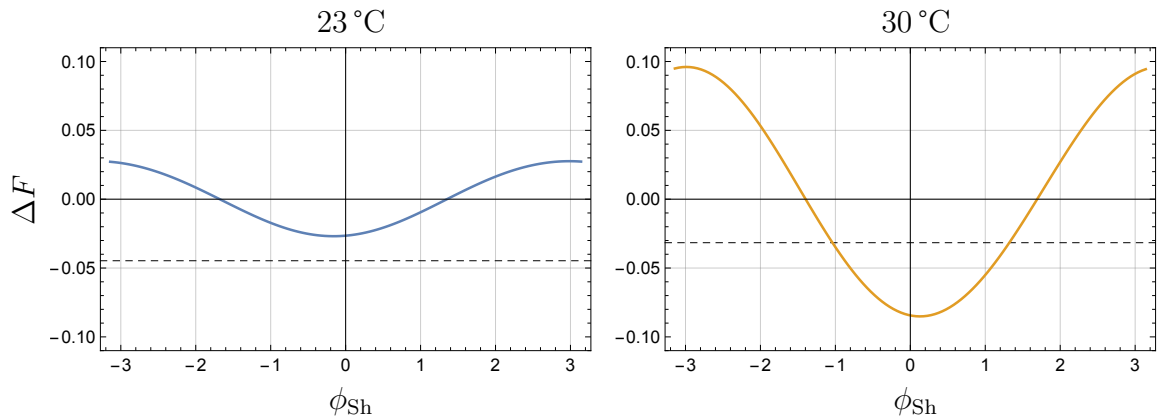
\begin{figure}[htb]
    \centering
	\begin{tikzpicture}
		\node (img23deg) {\includegraphics[scale=0.6]{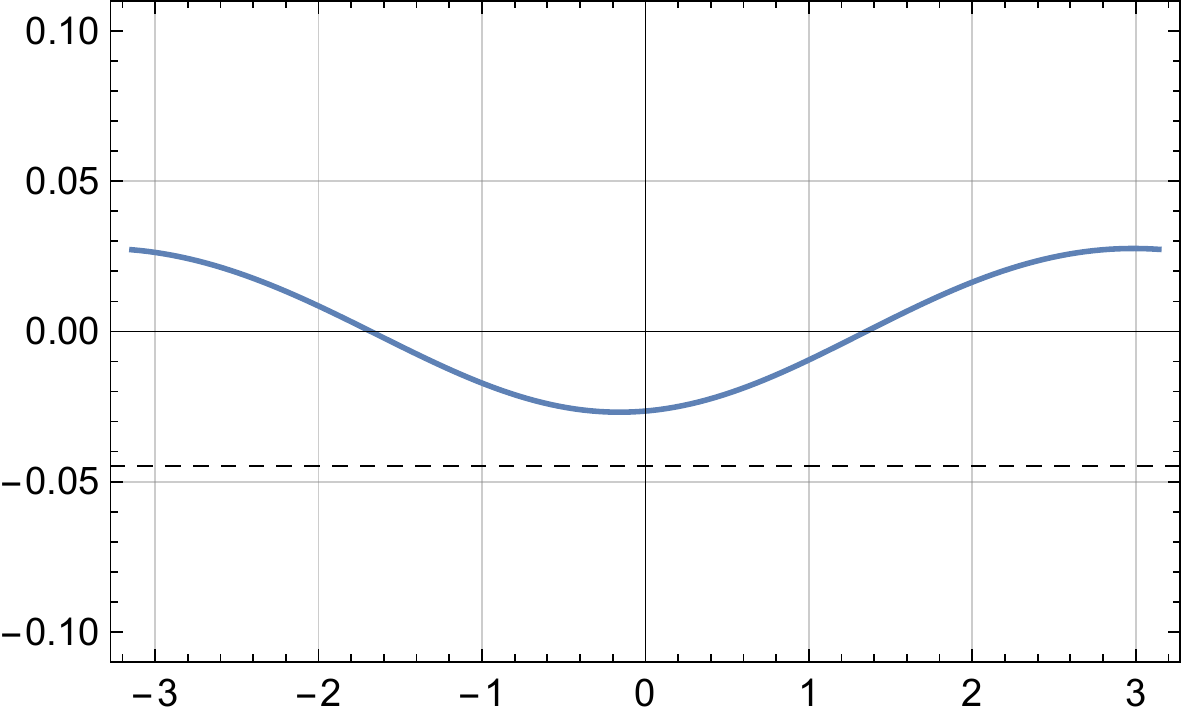}};
		\node[right=0 of img23deg] (img30deg) {\includegraphics[scale=0.6]{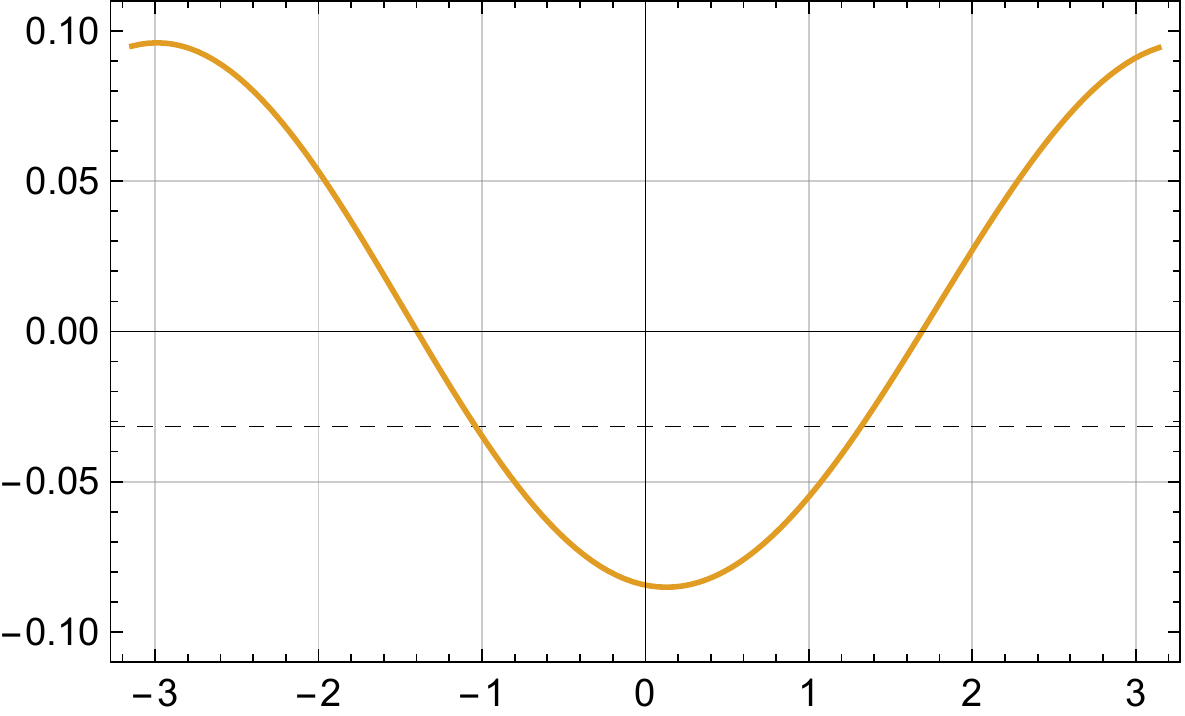}};
		\node[left=0 of img23deg, rotate=90, anchor=center,yshift=0.2cm,xshift=0.15cm] {$\Delta F$};
		\node[above=0 of img23deg, anchor=center, yshift=0.2cm, xshift=0.4cm]	{\SI{23}{\degreeCelsius}};
		\node[above=0 of img30deg, anchor=center, yshift=0.2cm, xshift=0.4cm]	{\SI{30}{\degreeCelsius}};
		\node[below=0 of img23deg, anchor=center, yshift=-0.2cm, xshift=0.4cm]	{$\phi_\mathrm{Sh}$};
		\node[below=0 of img30deg, anchor=center, yshift=-0.2cm, xshift=0.4cm]	{$\phi_\mathrm{Sh}$};
	\end{tikzpicture}
	\caption{Influence of residual light passing through closed shutters on the Peres parameter $F$. The deviation from the expected value $\Delta F = F-1$ is shown in dependence of the phase shift $\phi_\mathrm{Sh}$ caused by passing through a closed shutter. The result for a housing temperature of \SI{23}{\degreeCelsius} is shown in blue (left) and for \SI{30}{\degreeCelsius} in orange (right). The dashed lines describe the experimentally measured deviation for each temperature.}
	\label{fig:resLight}
\end{figure}

We estimated the shutter transmissivity based on the measured background signal $P_0$. For this calculation we assumed $P_i = \tilde P_0 + P_\mathrm{in} T_i$, i.e. neglecting the contribution of $\tau$ compared to $T_i$.
Furthermore we assumed that the shutter-independent background is only given by the dark voltage of the detectors $\tilde P_0 = P_\mathrm{dark}$.
Any additional shutter-independent background would further decrease the shutter transmissivity and, therefore, the result would act as an upper bound. Applying these assumptions to \eref{eqn:P0_resLight}, results in 
\begin{equation}
    \tau = \frac{P_0-\tilde P_0}{(\bar P_\mathrm{A} + \bar P_\mathrm{B} + \bar P_\mathrm{C} + 2\sqrt{\bar P_\mathrm{A} \bar P_\mathrm{B}} \gamma + 2\sqrt{\bar P_\mathrm{A} \bar P_\mathrm{C}} \beta + 2\sqrt{\bar P_\mathrm{B} \bar P_\mathrm{C}} \alpha)},
\end{equation}
where $\bar P_i = P_i - \tilde P_0$, describing the experimentally measured single path terms $P_i$ with the shutter-independent background $\tilde P_0$ already subtracted. For the set of normalized interferences $\alpha$, $\beta$ and $\gamma$, we used the corrected points in phase space. Evaluation of this equation for each measured cycle gives an average shutter transmissivity of 
\begin{equation}
    \mean{\tau_{\SI{23}{\degreeCelsius}}} = \num{2.20(2)E-4} \hspace{1em} \text{and} \hspace{1em} \mean{\tau_{\SI{30}{\degreeCelsius}}} = \num{2.707(4)E-4},
\end{equation}
with the uncertainty being the standard error of mean resulting from variations of the corrected points in phase space, as well as variations of the measured single path and background signals during the experiment. Applying these shutter transmissivities to our model results in the data shown in \cref{fig:resLight} for each housing temperature.
The deviation of the Peres parameter from its expected value can be negative as well as positive and strongly depends on the phase shift $\phi_\mathrm{Sh}$ of the residual light. The fact that the bias is stronger for the $\SI{30}{\degreeCelsius}$ measurement is not due to the slightly larger shutter transmissivity, but due to its point in phase space causing an increased sensitivity to the effect.
Due to the oscillations in $\Delta F$ with the unknown phase $\phi_{\mathrm{Sh}}$, we cannot reliably estimate a specific figure for the impact of the non-zero shutter transmissivity on our measured results, but can only provide a range given by its extrema. This range is wide enough to contain our experimental result ($\SI{30}{\degreeCelsius}$) or at least in the same order of magnitude as the observed deviation ($\SI{23}{\degreeCelsius}$).

\subsection{Conclusion}
The estimated effect of the various imperfections discussed and modelled in this work on the Peres parameter $F$ is summarized in \cref{tab:PeresSimRes}.
\begin{table}
	\caption{\label{tab:PeresSimRes}The calculated deviation of the Peres parameter $F$ from $F=1$ for different models and each of the two measurements at a housing temperature of \SI{23}{\degreeCelsius} and \SI{30}{\degreeCelsius}, respectively. Both fluctuation models only provide an upper bound for their effect and have a lower bound of 0. The residual light lists the deviation as an upper and lower bound, which corresponds to a phase shift of the transmitted light of $\phi_\mathrm{sh} \approx \pm\pi$ and $\phi_\mathrm{sh} \approx 0$, respectively. The last two rows list the total as an upper and lower bound and the experimentally measured deviation for each measurement.}
	\begin{indented}
    \lineup
    \item[] 
	\begin{tabular}{llll}
	\toprule
	& \multicolumn{2}{c}{$F-1$}	&  \\ \cmidrule{2-3}
	\tcellc{Model} & \tcellc{\SI{23}{\degreeCelsius}}	& \tcellc{\SI{30}{\degreeCelsius}}	& \tcellc{Notes} \\
	\midrule
	Detector Nonlinearity   & $\phle     \num{-4.7E-5}$ & $\phle     \num{-9.4E-5}$  & \\ \addlinespace
	Power Fluctuations      & $< \phm    \num{ 4.1E-4}$ & $< \phm    \num{ 1.4E-4}$  & \\ \addlinespace
	Phase Fluctuations      & $< \phm    \num{ 1.8E-4}$ & $>         \num{-8.1E-4}$  & \\ \addlinespace
	Interference Contrast   & $\phle     \num{-2.5E-6}$ & $\phle     \num{-4.2E-6}$  & \\ \addlinespace
	Crosstalk Phase         & $\phle \num{-6.4E-3}$ & $\phle\phm\num{ 6.6E-3}$  & \\ \addlinespace
	Residual Light  & \twoline{l}{$<\phm \num{2.8E-2}$}{$> \num{-2.7E-2}$}
	                & \twoline{l}{$<\phm \num{9.6E-2}$}{$> \num{-8.7E-2}$}
	                & \twoline{l}{$\phi_\mathrm{Sh} \approx \pm \pi$}{$\phi_\mathrm{Sh} \approx 0$}\\
	\midrule
	total   		& \twoline{l}{$<\phm \num{2.2E-2}$}{$> \num{-3.3E-2}$}	& \twoline{l}{$<\ \num{10.2E-2}$}{$>  \num{-8.1E-2}$} & \\
	\midrule
	measured		& $\phle \num{-4.47(4)E-2}$	& $\phle \num{-3.16(9)E-2}$	& \\
	\bottomrule
	\end{tabular}
    \end{indented}
\end{table}
The results suggest that the only imperfection resulting in deviations of the same order of magnitude as experimentally measured is the residual light transmission through closed shutters.
While in the case of the \SI{30}{\degreeCelsius} measurement this effect could have caused the entire deviation, for the \SI{23}{\degreeCelsius} measurement, it falls somewhat short of the observed deviation, even if all other considered effects are included. One has to keep in mind, though, that we have not taken into consideration the interplay of multiple error sources and that each model rests upon certain assumptions and simplifications. All other considered potential systematic error sources lead to effects at least one order of magnitude below the measured deviation. The different deviations for both housing temperatures show that the effect of each error source can be minimised by choosing an optimal set of phase differences.

In order to narrow down the range resulting from the residual light model, phase modulators in each of the paths would be necessary. 
This could be achieved by placing three heaters on the surface of the chip in between both splitters \cite{Flamini2015,Chaboyer2017,Dyakonov2018}. Arranging them in a similar triangle as the paths allows to create temperature gradients between each of the paths and with that tuning of the phase differences between them. Using phase modulators to sweep the phase of each path would allow determining $\phi_\mathrm{Sh}$ and thus increase the accuracy of this analysis. Alternatively, one could cancel out the interference effect (relative magnitude $\sqrt{\tau}$) by scrambling the phase in all closed channels, which would reduce the effect of residual light transmission (to order $\tau$). Even though the actual value of $\phi_{\mathrm{Sh}}$ and, with it, the actual impact on $F$ cannot be determined from our experiment, these results clearly show that shutters with a low residual transmission in the closed-state and, ideally, full phase control are critical for any interferometric test of hypercomplex quantum mechanics.
%
\ack
G.W.,  R.K., and S.G. acknowledge support from the Austrian Science Fund (FWF Projects No. I2562, M1849, and P30459). A.S. thanks the Deutsche Forschungsgemeinschaft (grants SZ 276/12-1, BL 574/13-1, SZ 276/21-1) and the EU Horizon2020 program (grants ErBeStA 800942, EPIQUS 899368). The authors gratefully acknowledge invaluable help of Mrs. Christiane Otto and Reinhard Geiss during the preparation of the samples.
\appendix
\section{Simulation details}

\subsection{Reconstruction of the point in phase space}
\label{sec:phaseSpaceReconstr}
The results of each systematic error model presented in this work depend on the point in phase space, where the Peres test is performed, i.e., the set of phase differences \{$\Delta\phi_\mathrm{BC}$, $\Delta\phi_\mathrm{CA}$, $\Delta\phi_\mathrm{AB}$\}. For conventional interference of complex amplitudes, by definition, the sum of this set must be zero (see \cref{eqn:PhaseSum}). 
However experimental imperfections, like the ones analysed in this work (cf. \cref{sec:SysErrors}), can lead to a deviation from zero and to seemingly non-physical phase differences. The goal here is to estimate the most probable original physical point in phase space, with phase differences summing to zero and therefore also $F=1$. This provides a starting point to estimate the effect of each individual experimental influence on the Peres parameter. The following section shows how we determined this point in detail.

\begin{figure}[htb]
	\begin{tikzpicture}
		\node (img23deg) {\includegraphics[scale=0.60]{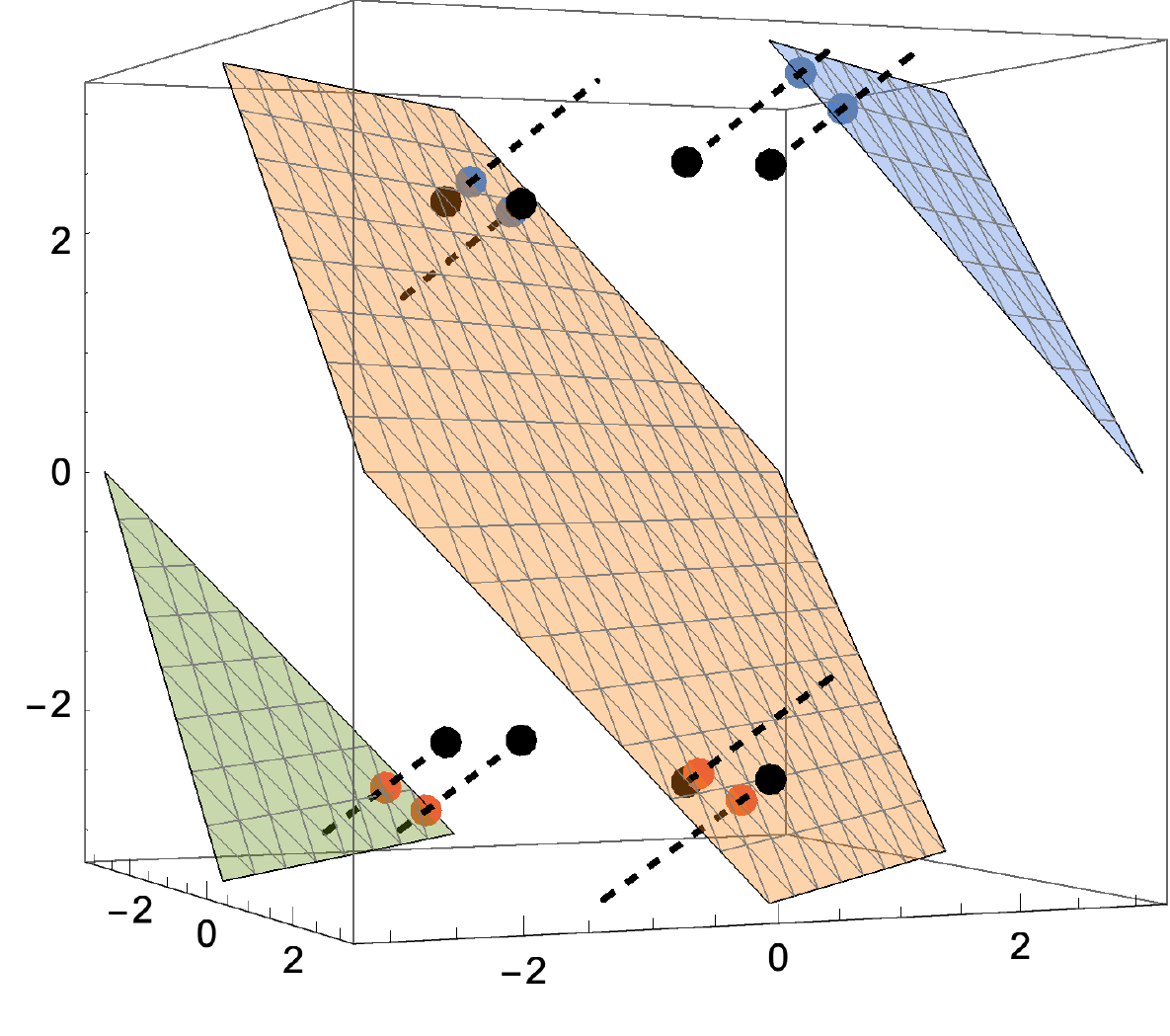}};
		\node[right=0 of img23deg, xshift=0.2cm] (img30deg) {\includegraphics[scale=0.60]{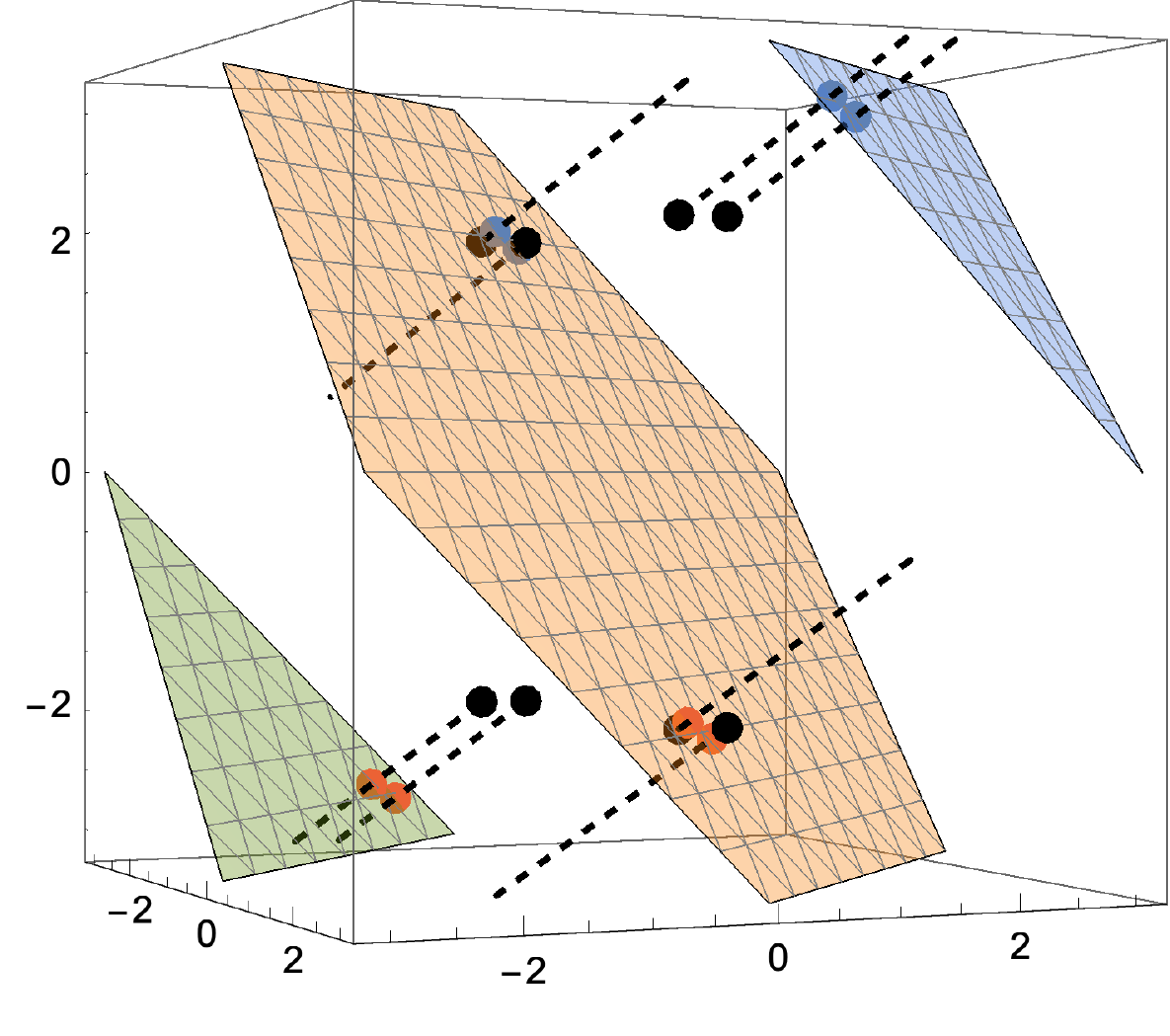}};
		\node[above=0 of img23deg, anchor=center, yshift=0.3cm, xshift=0.4cm]	{\SI{23}{\degreeCelsius}};
		\node[above=0 of img30deg, anchor=center, yshift=0.3cm, xshift=0.4cm]	{\SI{30}{\degreeCelsius}};
		\node[below=0 of img23deg, anchor=center, yshift=0.2cm, xshift=1.25cm]	{$\displaystyle\Delta\phi_\mathrm{CA}$};
		\node[below=0 of img30deg, anchor=center, yshift=0.2cm, xshift=1.25cm]	{$\displaystyle\Delta\phi_\mathrm{CA}$};
		\node[below=0 of img23deg, anchor=center, yshift=0.4cm, xshift=-2.65cm, rotate=-20]	{$\displaystyle\Delta\phi_\mathrm{BC}$};
		\node[below=0 of img30deg, anchor=center, yshift=0.4cm, xshift=-2.65cm, rotate=-20]	{$\displaystyle\Delta\phi_\mathrm{BC}$};
		\node[left=0 of img23deg, anchor=center, yshift=0.25cm, xshift=0.15cm, rotate=90]	{$\displaystyle\Delta\phi_\mathrm{AB}$};
		\node[left=0 of img30deg, anchor=center, yshift=0.25cm, xshift=0.15cm, rotate=90]	{$\displaystyle\Delta\phi_\mathrm{AB}$};
	\end{tikzpicture}
	\caption{
	\label{fig:phaseReconstr}Phase space of a three-path interferometer, which is defined by its three phase differences. The black points are calculated from the measurement data averaged over all cycles}. The planes describe the relation $\Delta\phi_\mathrm{BC} + \Delta\phi_\mathrm{CA} + \Delta\phi_\mathrm{AB} = 2 \pi n$ with $n \in \mathbb{Z}$. The case $n = 0$ is shown in orange, $n=1$ in blue and $n=-1$ in green. The orange and blue points represent the closest points on the plane to the measured points (black). The dashed lines connect each measured point with its point on the plain and indicate the direction of projection.
\end{figure}

The phase space is three-dimensional, as shown in \cref{fig:phaseReconstr}. The physically allowed phases $\Delta\phi_\mathrm{BC} + \Delta\phi_\mathrm{CA} + \Delta\phi_\mathrm{AB} = 2 \pi n$, with $n \in \mathbb{Z}$, form a periodic set of planes in phase space. From the measured set of phase differences we find the closest (using Euclidean distance) physical point in phase space by normal projection onto the planes. Due to the sign ambiguity in the cosine function (e.g. $\alpha = \cos(\Delta\phi_\mathrm{BC})$), the measured set $\{\alpha,\beta,\gamma\}$ corresponds to eight points. As a global sign flip of all phase differences has no effect on any of the models considered, this reduces to two sets of four inequivalent points, whose projections are shown in orange and blue, respectively. We therefore only consider one of the two sets e.g. the one depicted in blue.
Within each set, two points project onto the $n=0$ plane and two points onto the $\pm 2 \pi$-planes. The latter ones are excluded, as they would require much larger shifts, most likely due to inconsistent sign assignments. From the two remaining points we chose the one closer to the $n=0$ plane for each temperature. This results in the corrected phase differences listed in \cref{tab:phaseSpaceReconstr}. Note that choosing the second-closest point to the $n=0$ plane gives results in the same order of magnitude. The only exception is the phase crosstalk at \SI{30}{\degreeCelsius}, where the point farther from the $n=0$ plane leads to no real solution for $\Delta\phi_\mathrm{DH}$ (see \cref{sec:crosstalk}).

\subsection{Power and phase fluctuations}
\label{sec:CalcPhaseFluc}
To simulate the effect of power and phase fluctuations on the Peres parameter (see \cref{sec:fluctuations}) the magnitude of the respective fluctuations has to be determined. The signal of a single open path $P_i$ is only influenced by fluctuations of the input power and we therefore used its standard deviation to describe this fluctuation. In the case of two open paths, the measured signal fluctuations are a superposition of input-power and phase fluctuations. 
To estimate the strength of the power fluctuations we calculated the standard deviation of the two-path signals based on the single-path power fluctuations using Gaussian error propagation. The signal of two open paths $i$ and $j$ is described by
\begin{equation}
    P_{ij} = P_i + P_j + 2 \sqrt{P_i P_j} \cos(\Delta\phi_{ij}) 
\end{equation}
and therefore its input-power fluctuations $\sigma_{P_{ij}^\mathrm{pow}}$ as:
\begin{equation}
    \fl	\sigma_{P_{ij}^\mathrm{pow}} = \sqrt{\left( \sqrt{\frac{P_j}{P_i}} \cos(\Delta \phi_{ij}) +1 \right)^2 \left( \sigma_{P_i} \right)^2 + \left( \sqrt{\frac{P_i}{P_j}} \cos(\Delta \phi_{ij}) +1 \right)^2 \left( \sigma_{P_j} \right)^2},
\end{equation}
where $\sigma_{P_i}$ and $\sigma_{P_j}$ are the standard deviation of the single-path signal for path $i$ and $j$ respectively. We assumed phase and power fluctuations to be independent of each other and therefore the measured standard deviation of a two-path term is described by
\begin{equation}
    \label{eqn:flucTwoPathTerm}
    \sigma_{P_{ij}} = \sqrt{(\sigma_{P_{ij}^\mathrm{pow}})^2 + (\sigma_{P_{ij}^\mathrm{ph}})^2}.
\end{equation}
With this equation we calculated the phase fluctuation $\sigma_{P_{ij}^\mathrm{ph}}$ using the experimentally measured fluctuation $\sigma_{P_{ij}}$ for the open paths $i$ and $j$.

\subsection{Measuring the interference contrast}
\label{sec:InterfContrMeasurement}
In order to measure the interference contrast independently from estimations of the magnitude of phase fluctuations, we reduced the interferometer to a two-path one by keeping one path closed. Sweeping the phase difference over a range including fully constructive or destructive interference allows to directly measure the interference contrast. 
By creating a vertical temperature gradient along the triangular path arrangement, only the phase differences between path B (tip of the triangular path arrangement) and path A or C (base of the triangle) can be tuned. We decided to closed path A and sweep the phase difference between path B and C. To tune this temperature gradient we heated the sample to about \SI{34.9}{\degreeCelsius} and shut off the heater afterwards. We measured all four shutter combinations of path B and C in random order about 70 times, during the thermalization of the setup with its surrounding.
To increase the time resolution we only took five successive measurements per combination. Each combination took about \SI{5.71}{\second} including \SI{5.16}{\second} for changing the shutter combination. Each cycle takes about \SI{23.25}{\second} with \SI{0.41}{\second} for logging additional parameters like input power and housing temperature.

\begin{figure}[htb]
	\begin{tikzpicture}
		\node (temp) {\includegraphics[scale=0.58]{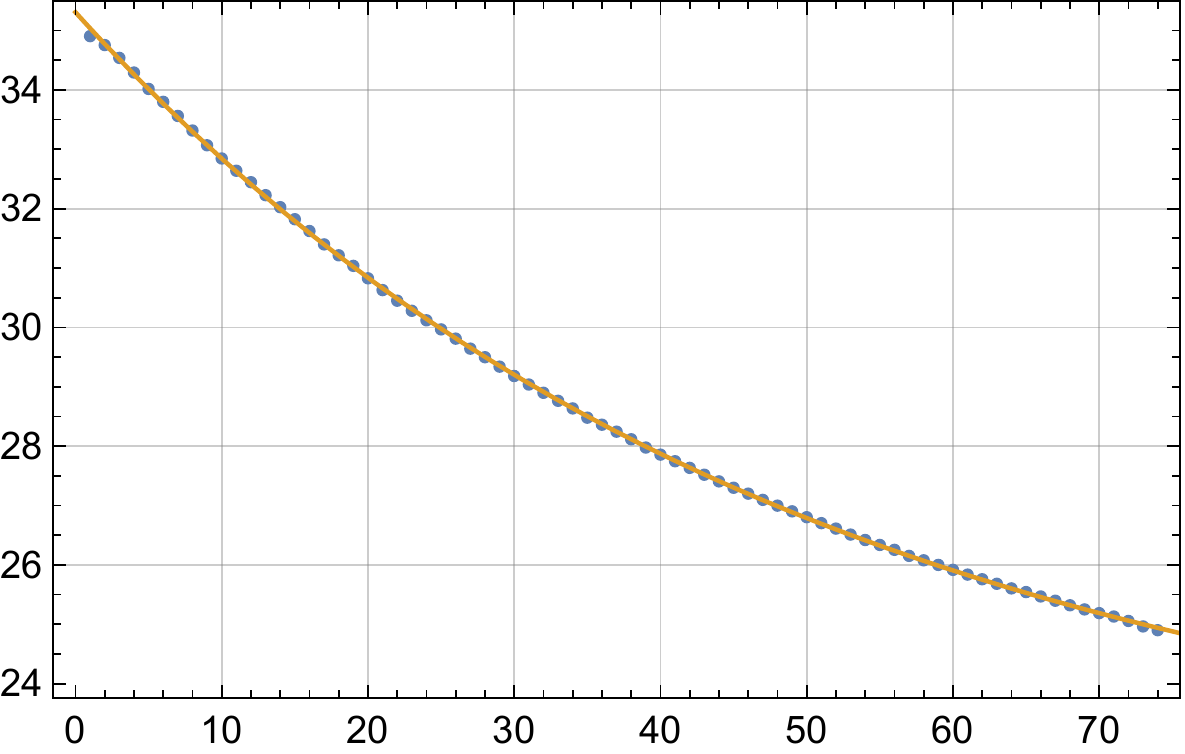}};
		\node[right=0 of temp, xshift=0.5cm] (interf) {\includegraphics[scale=0.59]{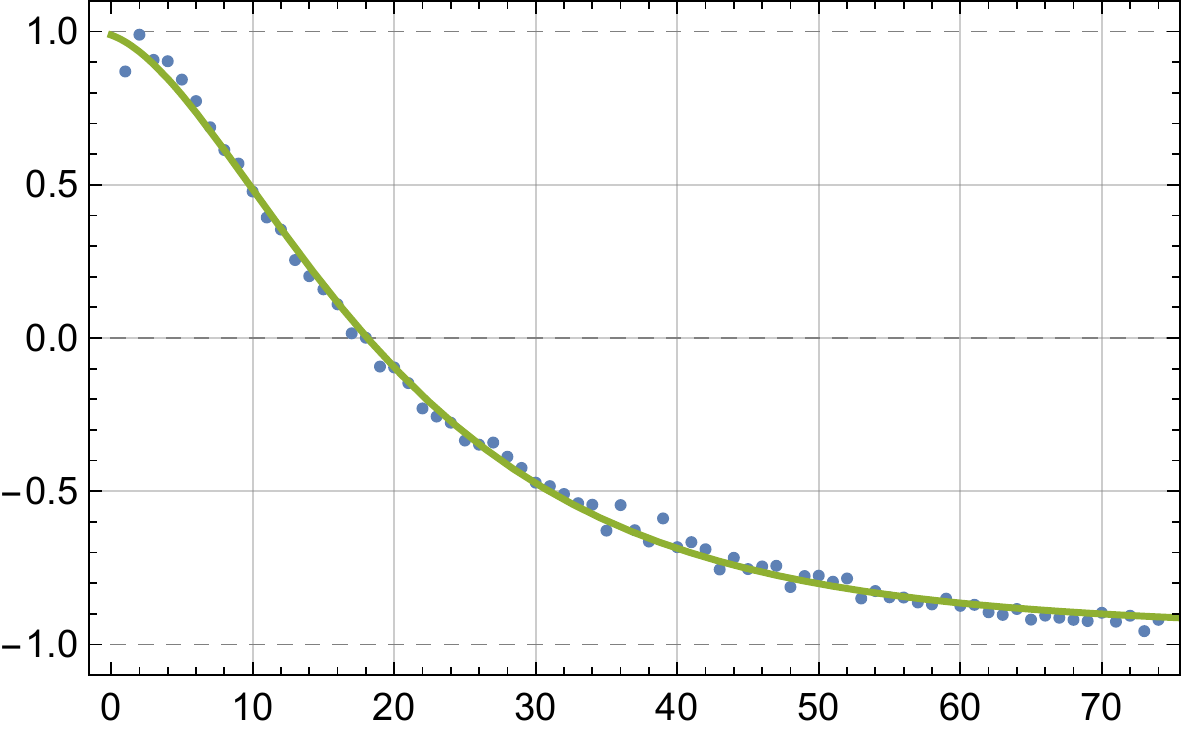}};
		\node[left=0 of temp,   rotate=90, anchor=center,yshift=0.2cm] {$T_\mathrm{h}$};
		\node[left=0 of interf, rotate=90, anchor=center,yshift=0.05cm] {$\alpha$};
		\node[below=0 of temp,  anchor=center, yshift=-0.2cm, xshift=0.4cm]	{$k$};
		\node[below=0 of interf,anchor=center, yshift=-0.2cm, xshift=0.4cm]	{$k$};
	\end{tikzpicture}
	\caption{Thermalization of the chip housing with its surroundings over multiple cycles $k$. In total the duration of the measurement was about \SI{28}{\minute}. The left plot shows the exponential drop of the housing temperature $T_\mathrm{h}$ during the thermalization and the solid orange line describes an exponential fit.
	The right plot shows the change of the normalized interference term $\alpha$ over the same time period.  The green solid line describes a fit using all parameters from the temperature fit, besides the thermalization coefficient, which is set as a free parameter.
	}
	\label{fig:interfContrast}
\end{figure}

The thermalization of the housing temperature $T_\mathrm{h}$ can be described by an exponential function
\begin{equation}
    T_\mathrm{h} = T_0 + \Delta T \exp (- \kappa k),
\end{equation}
where $k$ is the cycle index, $T_0$ the room temperature, $\Delta T$ the range of the temperature change and $\kappa$ the thermalization coefficient. The recorded temperature is shown together with an exponential fit ($T_0 = \SI{22.04(3)}{\degreeCelsius}$, $\Delta T = \SI{13.269(28)}{\degreeCelsius}$ and $\kappa = \SI{0.02055(10)}{\per cycle}$) in \cref{fig:interfContrast}.
This description fits the data quite well and therefore the influence on the normalised interference can be modelled by
\begin{equation}
    \alpha = C_\alpha \cos (\Delta\phi_0 + \eta \cdot\Delta T \exp (- \kappa k)), 
    \label{eqn:interfContr_fit}
\end{equation}
where $C_\alpha$ describes the interference contrast, $\Delta\phi_0$ the phase difference at room temperature and $\eta$ the change of the phase difference per temperature change. This model, with the parameters $\kappa = \SI{0.0415(13)}{\per cycle}$, $\Delta\phi_0 = \SI{3.44(7)}{\radian}$, $\eta = \SI{0.203(9)}{\radian\per\degreeCelsius}$ and $C_\alpha = \num{1.000(29)}$, fits the measured data very well as shown in \cref{fig:interfContrast}.
The fit parameter of $C_\alpha$ shows that the interference contrast of these two paths is, within the experimental precision of this measurement, quite high.

Note that the parameters of the temperature fit were not used for fitting the normalised interference, due the position of the temperature sensor. It is mounted on the outside of the housing, between the insulation and the housing. Therefore the measured housing temperature does not perfectly describe the change of the temperature gradient and with it the change of the phase difference.
Also the phase change per temperature $\eta$, resulting from the fit of the left plot in \cref{fig:interfContrast}, can not be compared with the one given in \cref{sec:Setup}, which described a fully thermalized situation and here we have a dynamic thermal process.

\subsection{Polarization}
\label{sec:polarization}
The three-dimensional coupling between the waveguides in the fused silica chip leads to polarization rotations. These changes in polarization depend on the orientation of the  adjacent waveguides in regards to the orientation of their elliptical waveguide profile. Therefore, the two waveguides at the base of the triangle are affected differently than the one at the tip. This effect leads to a superposition of light of different polarizations in the output ports of the chip. Without a polarizer in front of the detector the combined signal of both orthogonal polarizations is measured. For example if only one path $i$ is open this can be described using the horizontal H and vertical V polarization basis as
\begin{equation}
    P_i = P_i^\mathrm{H} + P_i^\mathrm{V}.
\end{equation}
In case of a two open paths $i$ and $j$ the same principle can be applied leading to 
\begin{equation}
\fl    P_{ij} = P_i^\mathrm{H} + P_j^\mathrm{H} + 2 \sqrt{P_i^\mathrm{H} P_j^\mathrm{H}} \cos(\Delta \phi_{ij}^\mathrm{H}) + P_i^\mathrm{V} + P_j^\mathrm{V} + 2 \sqrt{P_i^\mathrm{V} P_j^\mathrm{V}} \cos(\Delta \phi_{ij}^\mathrm{V}).
\end{equation}
A polarizer in front of the detector can be set to filter all H-components resulting in $F=1$, which is trivial to show using \cref{eqn:PeresParameter} and \cref{eqn:gamma}-\eref{eqn:beta}.

Without a polarizer the resulting equation for $F$ is quite long and does not reduce to 1. This can be shown for the case of equal powers for both polarizations $P_i^\mathrm{H} = P_i^\mathrm{V} = P_i/2$. 
This reduces $P_{ij}$ to
\begin{equation}
    P_{ij} = P_i + P_j + \sqrt{P_i P_j} (\cos(\Delta \phi_{ij}^\mathrm{H}) + \cos(\Delta \phi_{ij}^\mathrm{V})).
\end{equation}
Calculating $F$ for this case results in
\begin{equation}
    \fl F = \frac{1}{2} \left( 1+ \alpha^\mathrm{H} \alpha^\mathrm{V} + \beta^\mathrm{H} \beta^\mathrm{V} + \gamma^\mathrm{H} \gamma^\mathrm{V} - \frac{1}{2} \sum_{i,j,k \in \{\mathrm{H},\mathrm{V}\}} \alpha^i \beta^j \gamma^k + \alpha^\mathrm{H} \beta^\mathrm{H} \gamma^\mathrm{H} + \alpha^\mathrm{V} \beta^\mathrm{V} \gamma^\mathrm{V} \right).
\end{equation}
It is easy to find normalised interference terms $\alpha^i$, $\beta^i$ and $\gamma^i$, which fulfill $F^i=1$ for each polarization $i$, but result in $F \neq 1$. This shows that a measurement of a combination of orthogonal polarizations can lead to a deviation in the Peres parameter. In contrast, by projecting to a single polarization, this effect can be avoided.

\section{Determination of crosstalk via Sorkin test}
\label{sec:Sorkin_Appendix}
To determine the phase crosstalk $\Delta\phi_\mathrm{DH}$ in our experiment we utilize the Sorkin test \cite{Sorkin1994, Sorkin1997}, usually applied to probe the presence of higher-order interferences, which would violate Born's rule \cite{Sinha2010,Hickmann2011,Park2012,Jin2017,Kauten2017,Cotter2017,Barnea2018,Pleinert2020}. We resorted to this test as it is unbiased by most error sources, except by nonlinearity in the setup and crosstalk between the modes. For a three-path interferometer, as used in our experiment, the output rate for all shutters open can be obtained via Born's rule as
\begin{equation}
    \label{eqn:PABCstandard}
    P_\mathrm{ABC} = \abs{\psi_\mathrm{A} + \psi_\mathrm{B} + \psi_\mathrm{C}}^2 = 
    P_\mathrm{AB} + P_\mathrm{BC} + P_\mathrm{CA} - P_\mathrm{A} - P_\mathrm{B
    } - P_\mathrm{C}.
\end{equation}
Any experimental deviation from this result can be subsumed in the rate $\epsilon$:
\begin{equation}
    \label{eqn:PABCmeas}
    \epsilon \defeq P_\mathrm{ABC} + P_\mathrm{A} + P_\mathrm{B} + P_\mathrm{C} - P_\mathrm{AB} - P_\mathrm{BC} - P_\mathrm{CA} 
\end{equation}
According to Sorkin, deviations from $\epsilon=0$ indicate higher-order interference and would violate Born's rule. To foster a comparison across different experimental platforms, a normalized version of this rate (called Sorkin parameter) is used. A common variant is the normalization of the mean of $\epsilon$ in regards to the mean sum of the absolute values of the pairwise interference terms:
\begin{equation}
    \label{eqn:SorkinParameter_norm}
    \mean{\kappa} \defeq \frac{\mean{\epsilon}}{\mean{\abs{P_\mathrm{AB} - P_\mathrm{A} - P_\mathrm{B}} + \abs{P_\mathrm{BC} - P_\mathrm{B} - P_\mathrm{C}} + \abs{P_\mathrm{AC} - P_\mathrm{A} - P_\mathrm{C}}}}
\end{equation}
This normalization makes $\kappa$ independent of input power and interference contrast.

\begin{figure}[htb]
	\begin{tikzpicture}
		\node (img1) {\includegraphics[scale=0.605]{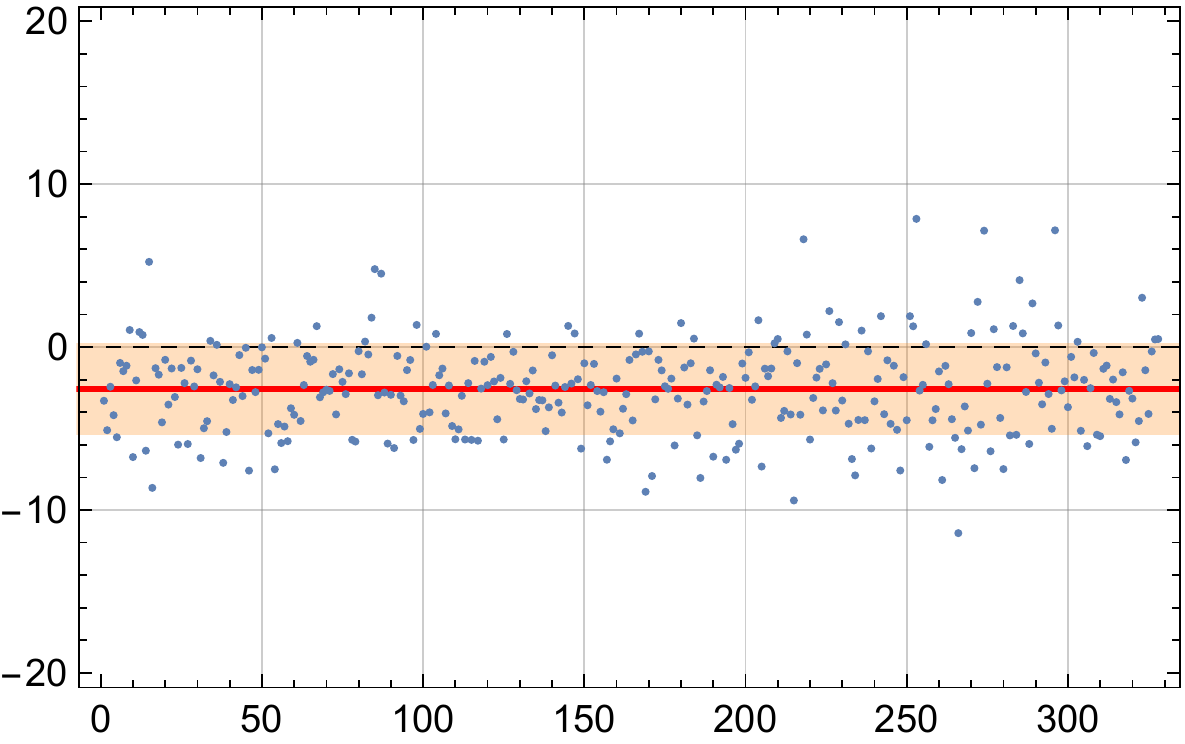}};
		\node[right	=0 of img1]	(img2) {\includegraphics[scale=0.605]{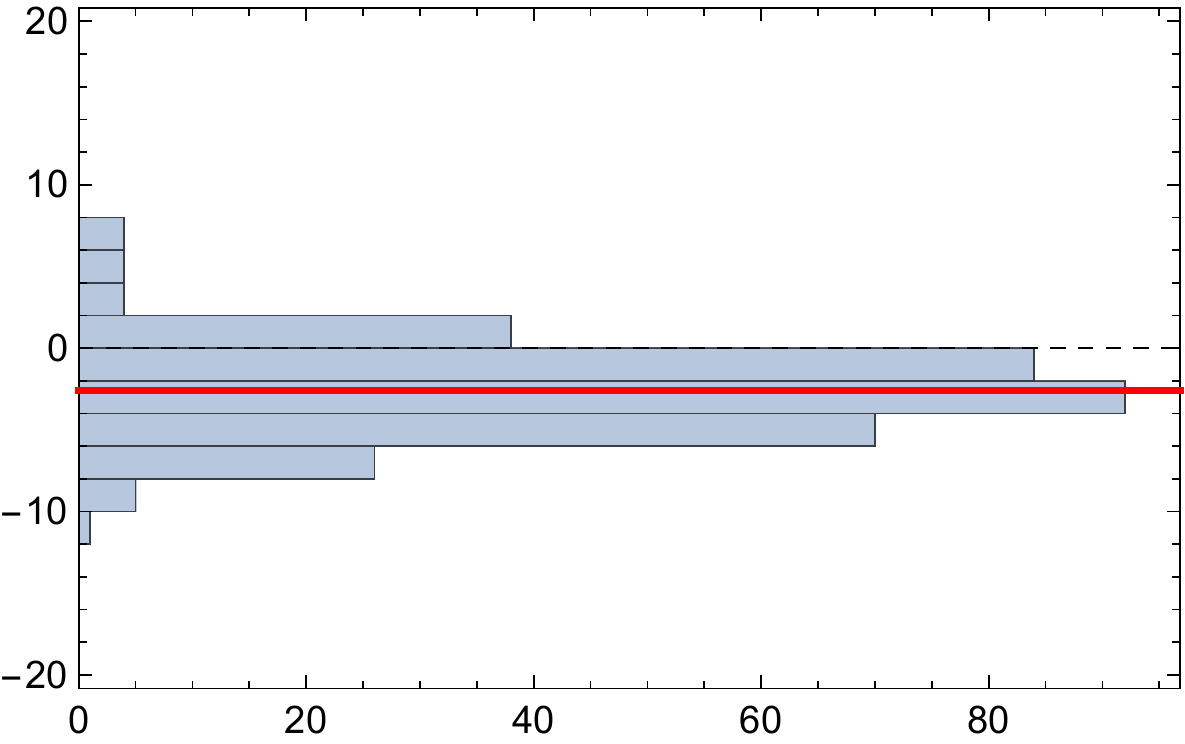}};
		\node[below =0 of img1] (img3) {\includegraphics[scale=0.605]{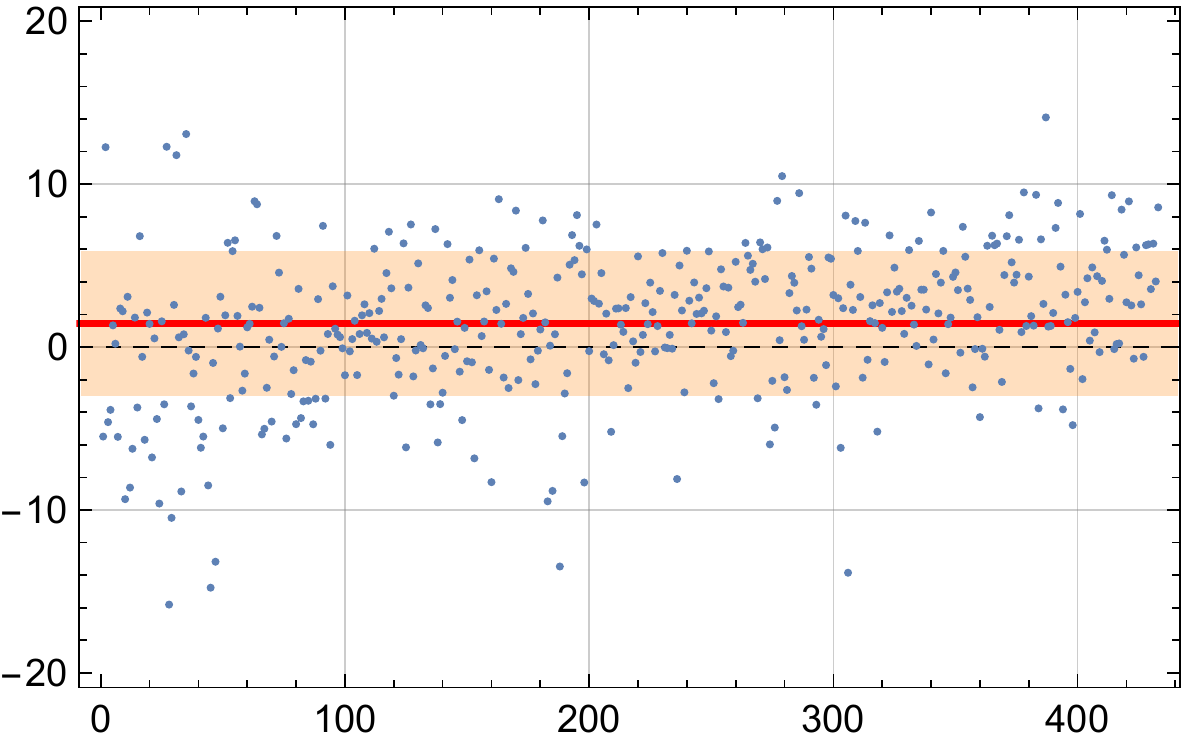}};
		\node[right	=0 of img3]	(img4) {\includegraphics[scale=0.605]{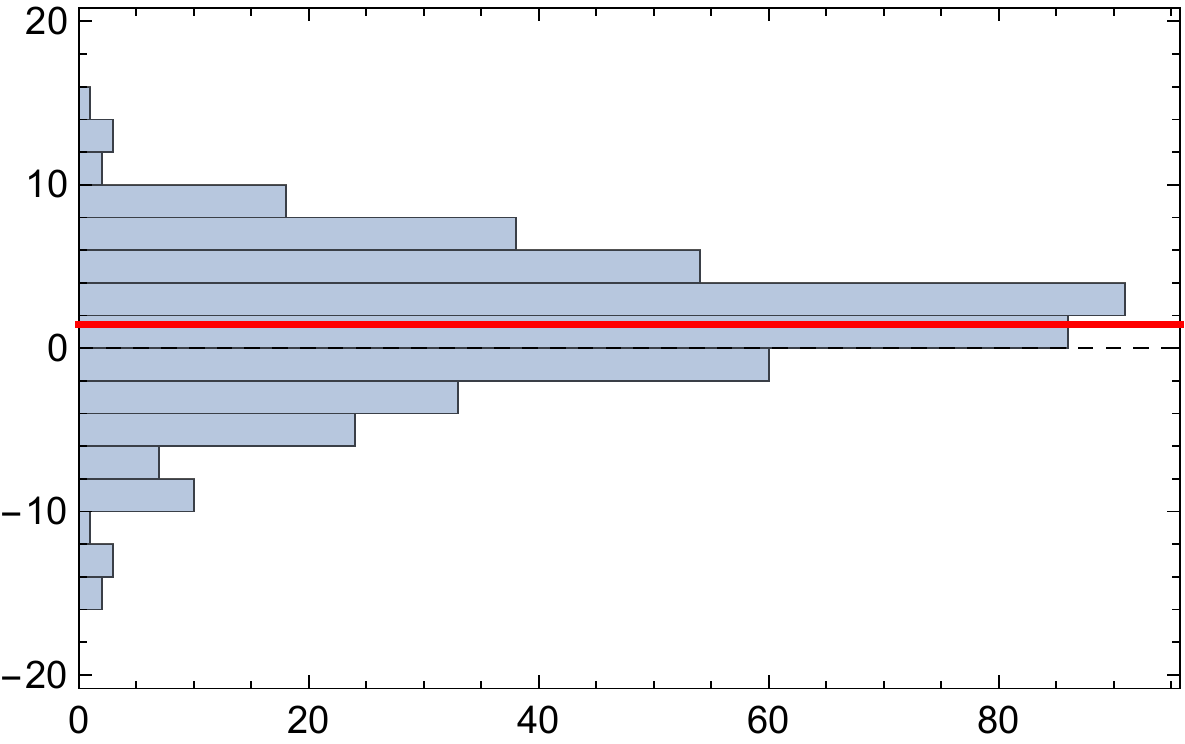}};
		\node[left=0 of img1, rotate=90, anchor=center,yshift=0.2cm,xshift=0.15cm] {$\epsilon$ (\si{\nano\watt})};
		\node[left=0 of img3, rotate=90, anchor=center,yshift=0.2cm,xshift=0.15cm] {$\epsilon$ (\si{\nano\watt})};
		\node[below=0 of img3, anchor=center, yshift=-0.2cm, xshift=0.4cm]	{$k$};
		\node[below=0 of img4, anchor=center, yshift=-0.2cm, xshift=0.4cm]	{Frequency};
		\node[above right=-0.75 of img1, anchor=north east, yshift=0.2cm, fill=white] {\SI{23}{\degreeCelsius}}; 
		\node[above right=-0.75 of img2, anchor=north east, yshift=0.2cm, fill=white] {\SI{23}{\degreeCelsius}}; 
		\node[above right=-0.75 of img3, anchor=north east, yshift=0.2cm, fill=white] {\SI{30}{\degreeCelsius}};
		\node[above right=-0.75 of img4, anchor=north east, yshift=0.2cm, fill=white] {\SI{30}{\degreeCelsius}};
	\end{tikzpicture}
	\caption{The rate $\epsilon$ (in blue) for multiple successive measurements, with index $k$, on the left and the corresponding histogram on the right. The plots of each row were measured at different housing temperature as stated in the top right corner of each plot. The red horizontal lines describe the mean values and the orange shaded bands represents one standard deviation. The expected value of $\epsilon=0$ is indicated by a black dashed line.
    }
	\label{fig:sorkin_param}
\end{figure}

Performing the Sorkin test with the same measurement data, supplemented by data of the rate $P_\mathrm{ABC}$, resulted in the data as shown in \cref{fig:sorkin_param} and the following mean values for each housing temperature:
\begin{eqnarray}
    \mean{\epsilon_{\SI{23}{\degreeCelsius}}} = \SI{-2.58(16)}{\nano\watt} \hspace{0.5em} \text{and} \hspace{0.5em}
    \mean{\epsilon_{\SI{30}{\degreeCelsius}}} = \SI{1.5(5)}{\nano\watt}\\
    \mean{\kappa_{\SI{23}{\degreeCelsius}}} = \num{-14.0(8)E-4} \hspace{0.5em} \text{and} \hspace{0.5em}
    \mean{\kappa_{\SI{30}{\degreeCelsius}}} = \num{11(4)E-4}.
\end{eqnarray}
The given error is the standard error of mean corrected by the autocorrelation according to \cite{Zhang2006}. Using this method for the error calculation, both housing temperatures show a significant deviation from the expected value $\epsilon = \kappa = 0$.
When comparing these values with the literature one has to keep in mind, that this test is usually performed with an interferometer tuned to fully constructive or destructive interference, where phase deviations have smaller effects. As this is not the case in our experiment, the systematic deviations from 0 are larger than they would be at the extremal points.

As higher-order interferences have been experimentally ruled out in other experiments for coherent light illumination to at least $\left|\kappa\right|<3\times10^{-5}$ \cite{Kauten2017}, any apparent higher-order interference appearing in the present experiment can be attributed to nonlinearity or crosstalk. We determined the nonlinearity to be negligible by independent measurements (see \cref{sec:DetectorNonLinearity}), such that effectively the Sorkin test turns into a measure of crosstalk. We can therefore associate systematic deviations from the predicted $\epsilon=0$ to cross-channel influences of shutter settings between the waveguides (see discussion in \cref{sec:crosstalk}). 
To calculate the phase crosstalk $\Delta\phi_\mathrm{DH}$ based on the measured $\epsilon$, we used the normalised interference terms modified by the phase crosstalk $\Delta\phi_\mathrm{DH}$ (as described in \cref{eqn:crosstalkNormInterf}) and inserted them into the corresponding two-path rates $P_{ij}$ in \cref{eqn:PABCmeas}. This results in the following equation connecting the phase crosstalk $\Delta\phi_\mathrm{DH}$ with the rate $\epsilon$:
\begin{equation}
    \fl    \epsilon = 2 P_\mathrm{in} \left(\sqrt{T_\mathrm{A} T_\mathrm{B}} \left( \gamma - \cos(\Delta\phi_\mathrm{AB} - \Delta\phi_\mathrm{DH})\right) + \sqrt{T_\mathrm{B} T_\mathrm{C}} \left( \alpha - \cos(\Delta\phi_\mathrm{BC} + \Delta\phi_\mathrm{DH}) \right) \right)
\end{equation}
For each temperature, this equation leads to two solutions for $\Delta\phi_\mathrm{DH}$, which differ by one order of magnitude. We chose the solution with the lower crosstalk strength (given in \cref{tab:Crosstalk}), as we expect it to be a small effect.

\section*{References}
\bibliography{main}

\end{document}